\documentclass[showpacs,preprintnumbers,amsmath,nofootinbib,amssymb,superscriptaddress]{revtex4}

\usepackage{amsmath,color}
\usepackage{amssymb}
\usepackage{bm}% bold math
\usepackage{graphicx}
\usepackage{multirow}
\usepackage[hidelinks]{hyperref}

\usepackage{textcomp}

\allowdisplaybreaks[1]

\newcommand{\nn}{\nonumber \\}
\newcommand{\beq}{\begin{eqnarray}}
\newcommand{\eeq}{\end{eqnarray}}
\newcommand{\lT}{\boldsymbol{l}_T}
\newcommand{\loT}{\boldsymbol{l}_{1T}}
\newcommand{\ltT}{\boldsymbol{l}_{2T}}

\newcommand{\Slash}[1]{{\ooalign{\hfil/\hfil\crcr$#1$}}}

\begin{document}

\preprint{YITP-19-78}

\title{\boldmath
Single-spin asymmetries at two loops
}

\author{Sanjin Beni\' c\footnote{On leave of absence from: Department of Physics, Faculty of Science, 
University of Zagreb, Bijeni\v cka c. 32, 10000 Zagreb, Croatia}}
\affiliation{Yukawa Institute for Theoretical Physics,
Kyoto University, Kyoto 606-8502, Japan}

\author{Yoshitaka Hatta}
\affiliation{Physics Department, Brookhaven National Laboratory, Upton, New York 11973, USA}

\author{Hsiang-nan Li}
\affiliation{Institute of Physics, Academia Sinica,
Taipei, Taiwan 11529, Republic of China}

\author{Dong-Jing Yang}
\affiliation{Department of Physics, National Taiwan Normal University, Taipei, Taiwan 10610, Republic of China}

\date{\today}

\begin{abstract}

We find a novel mechanism for generating transverse single-spin asymmetry (SSA) in semi-inclusive deep inelastic 
scattering, distinct from the known ones which involve the Sivers and Collins functions, 
or their collinear twist-three counterparts. It is demonstrated that a phase needed for SSA can
be produced purely within a parton-level cross section starting at two loops.
We identify the complete set of two-loop diagrams for SSA, and discuss their gauge invariance and collinear 
factorization which features the $g_T$ distribution function.
In the $k_T$ factorization framework, many more sources for SSA exist,
and contributions from all possible two-parton transverse-momentum-dependent parton distribution functions
are presented up to two loops and twist three.

\end{abstract}

%\pacs{12.38.-t, 12.38.Bx, 12.38.Gc, 14.20.Dh}

%
%
%

\maketitle

%
%
%
%--------+---------+---------+---------+---------+---------+---------+---------+

\section{INTRODUCTION}

A study of  single-spin asymmetry (SSA) in  processes
involving a transversely polarized nucleon is crucial for exploring the
three-dimensional nucleon structure. Significant experimental signals of SSA
have been observed in hadron production \cite{B76,A91}, which amount
up to order of ten or more percent of unpolarized cross sections.
Data on pion production have been very consistent, showing asymmetries up to pion transverse momenta of 
several GeV \cite{Adare:2013ekj,Adamczyk:2012xd}. Despite decades of efforts, the origin of such significant 
asymmetries is not yet fully understood, due in part to large theoretical and experimental uncertainties. The 
future Electron-Ion Collider is expected to deliver very precise measurements, that will impose strong 
constraints on various theoretical approaches.

From the theoretical point of view, understanding SSA is a quest for a `phase'.
One is interested in the part of a cross section which depends linearly on the transverse spin vector
${\boldsymbol S}_T$ of a nucleon. The spin vector usually comes with a factor $i$, so to make the cross
section real, one has to find another factor $i$ from involved diagrams.
The first such attempt was made by Kane, Pumplin and Repko \cite{KPR78}, who calculated the SSA
for single hadron (pion) production from quark-quark scattering diagrams with a transversely polarized
quark. They found that nonvanishing SSA for high $p_T$ reactions is proportional to a current quark mass.
Although their calculation does not explain the measured large SSA, the observation 
with the result being proportional to a quark mass indicates that SSA is a twist-three effect in
perturbative QCD. Subsequently, Efremov and Teryaev pointed out that nonvanishing SSA could be obtained
as one goes beyond the leading power \cite{Efremov:1981sh,Efremov:1983eb,Efremov:1984ip,Ratcliffe:1985mp}. 
It is by now well known that sizable SSA can be generated through the combined effect of nonperturbative 
twist-three distributions of a nucleon, called the Efremov-Teryaev-Qiu-Sterman (ETQS) function 
\cite{Efremov:1981sh,Efremov:1983eb,Qiu:1991pp,Qiu:1998ia}, and the pole part of a propagator which 
provides the required phase. In this picture, the smallness of a current quark mass is no longer an issue, 
since the relevant mass scale is a nucleon mass. A similar twist-three effect has been implemented into 
fragmentation functions as an alternative source of SSA 
\cite{Kang:2010zzb,Metz:2012ct,Kanazawa:2013uia}.

SSA has been also studied extensively in the $k_T$ factorization framework. Parton transverse momenta
are incorporated either in transverse-momentum-dependent (TMD) parton distribution functions  (PDFs) or in
TMD fragmentation functions (FFs). The former is the Sivers function \cite{S90,ABM95}, which describes the
spin-orbit correlation of partons inside a transversely polarized nucleon. The required phase arises from 
the pole of a propagator for Wilson lines. For the latter, the Collins
function \cite{C93,Collins:1993kq,ACY97} governs the fragmentation of a polarized quark, in which the
phase comes from final state interactions.

In this paper we will investigate the source of phases starting from a
parton-level cross section up to two loops, taking the polarized semi-inclusive
deeply inelastic scattering (SIDIS) as an example. On-shell internal particles in certain two-loop diagrams 
produce phases from different leading regions of particle momenta.
The phase is then absorbed into the relevant piece in the factorization theorem for each leading region. 
In addition to the known Sivers (or ETQS) and Collins
mechanisms, which are associated with the collinear regions of initial state
and final state partons, respectively, we find a novel source of phases
which goes into a hard kernel. The corresponding factorization
formula contains the $g_T$ distribution function for a polarized nucleon, and the standard twist-2 FF for a
final state hadron. Our result is reminiscent of the observations 
in \cite{Ma:2008gm,Ma:2008cj,Metz:2006pe,Afanasev:2007ii}: the authors of \cite{Ma:2008gm,Ma:2008cj} studied the 
same set of two-loop diagrams as proposed in this work, but for a transversely polarized quark target.
The asymmetry is thus proportional to a current quark mass, and only factorizations into the known mechanisms 
(Sivers, ETQS,...) were examined. In \cite{Metz:2006pe,Afanasev:2007ii}, the authors found 
that multi-photon exchange between the leptonic and hadronic parts of inclusive 
deep-inelastic lepton-hadron scattering causes SSA.
The two-photon-exchange diagrams considered in \cite{Metz:2006pe}
has the same topology as our diagrams, but it turned out that their final formula does not contain 
the $g_T$ distribution function \cite{Schlegel:2012ve}.

Once we are allowed to go to higher orders in a hard cross section, more
twist-3 TMD PDFs and FFs from various spin projectors can contribute
to SSA, resulting in abundant phenomenology to be explored. We will derive a complete
set of subleading contributions to transverse SSA at
two-parton twist-three accuracy in SIDIS up to two loops. Note that
the proof of the factorization theorem at the twist-three level is highly
nontrivial. Here we will adopt the twist-three factorization as a working
hypothesis \cite{Bacchetta:2008xw,Vogelsang:2009pj,Song:2010pf,Kang:2012ns,Yoshida:2016tfh,Chen:2017lvx}, and leave its rigorous
proof to future projects.

This paper is organized as follows. In Section II, we present the general formalism for SSA in the collinear factorization and check the QED and QCD gauge invariance. In Section III, the complete set of two-loop diagrams that should be included into the hard kernel introduced in Section II is identified. We analyze the various infrared divergences in the considered diagrams, and discuss how to handle these divergences in the collinear and $k_T$ factorizations in Section IV. A source of phase, which cannot be ascribed to the known mechanisms of SSA, will be highlighted. It thus represents a new contribution to SSA, and is our main result. Section V is the conclusion.

\section{Semi-inclusive deep inelastic scattering}

\begin{figure}[!]
\includegraphics[scale=0.5]{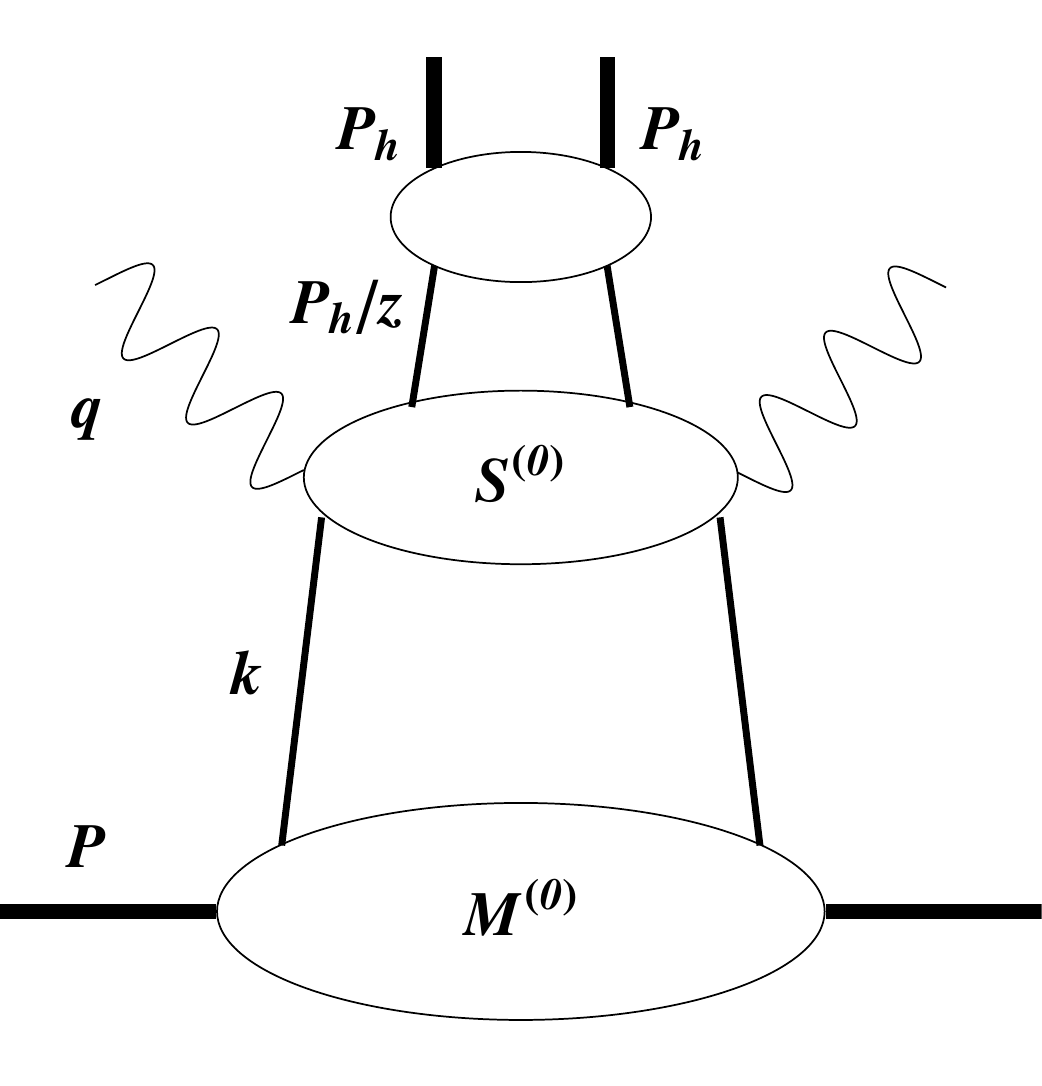}
\hspace{1.0cm}
\includegraphics[scale=0.5]{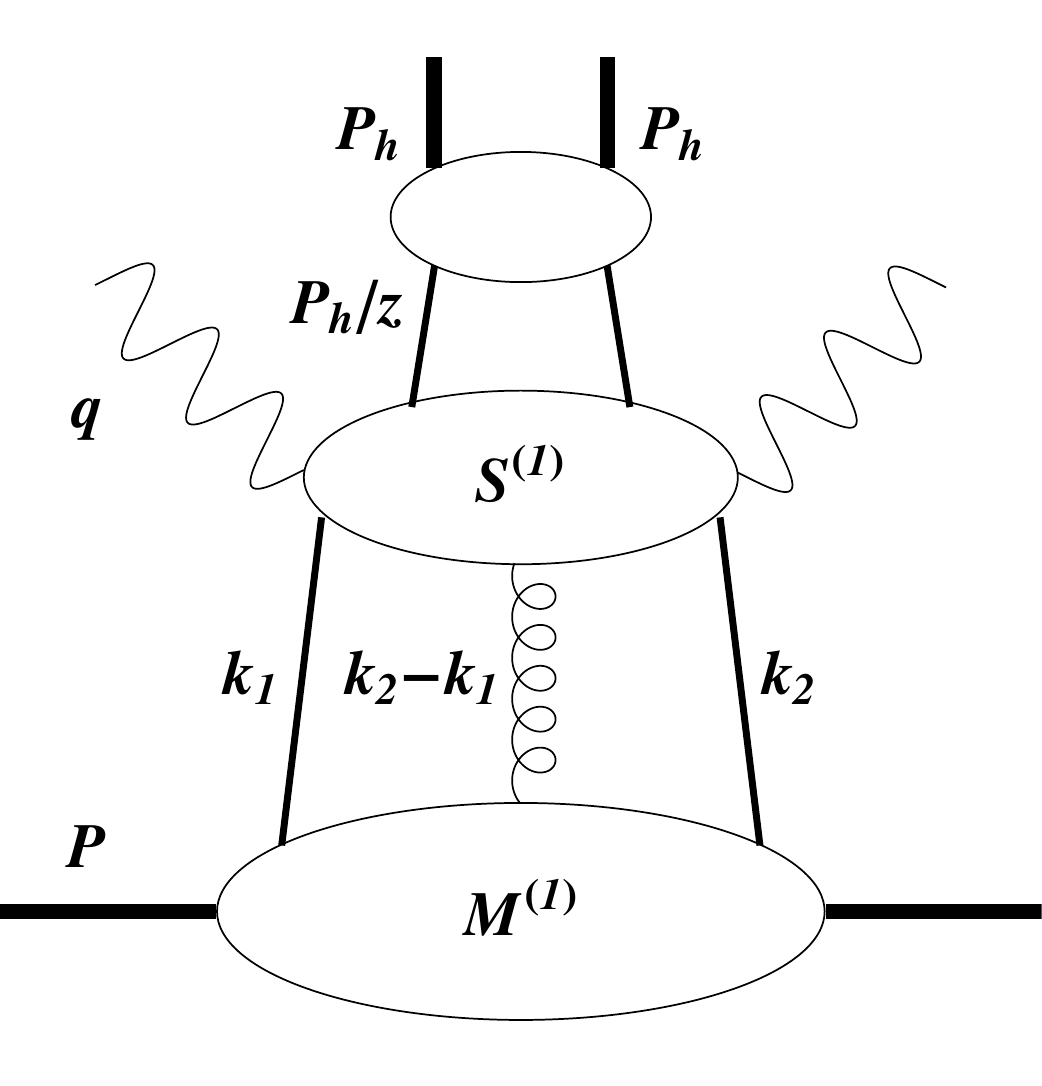}
\caption{Generic diagrams contributing to SSA in SIDIS, and a graphical representation of the two terms in Eq.~(\ref{master}).
} \label{dia}
\end{figure}

In this section we start with a general discussion of SSA in SIDIS in the collinear factorization framework 
mostly following the notations of \cite{Eguchi:2006mc} (see also \cite{Ratcliffe:1985mp,Qiu:1998ia}).
The spin-dependent part of the $e(l)p(P)\to e(l')h(P_h)X$ cross section is given by
\beq
d\sigma = \frac{1}{2S_{ep}}\frac{d^3P_h}{(2\pi)^32E_h}\frac{d^3 l'}{(2\pi)^32E_{l'}} \frac{e^4}{(Q^2)^2} L^{\mu\nu}W_{\mu\nu} ,
\eeq
where $S_{ep}\equiv (l+P)^2$, $Q^2\equiv -q^2=-(l-l')^2$, $L^{\mu\nu}=2(l^\mu l'^\nu + l^\nu l'^\mu)-g^{\mu\nu}Q^2$ is the leptonic tensor, $W^{\mu\nu}$ is the hadronic tensor, and $\nu$ and $\mu$ are  the polarization indices of the virtual photon in the amplitude and the complex-conjugate amplitude, respectively. The Bjorken variable is denoted as $x_B=Q^2/(2P\cdot q)$.  We work in the so-called hadron frame, where the virtual photon and the proton move in the $z$ direction with
\beq
q^\mu=(0,0,0,-Q), \qquad P^\mu=\left(\frac{Q}{2x_B},0,0, \frac{Q}{2x_B} \right).
\eeq
The incoming and outgoing leptons have the momenta
\beq
l^\mu = \frac{Q}{2} (\cosh\psi ,\sinh\psi \cos \phi, \sinh\psi \sin \phi, -1), \qquad l'^\mu = \frac{Q}{2} (\cosh\psi ,\sinh\psi \cos \phi, \sinh\psi \sin \phi, 1), \label{use}
\eeq
where $\phi$ is the azimuthal angle relative to the $z$ axis, and
\beq
\cosh\psi \equiv \frac{2x_BS_{eq}}{Q^2}-1 = \frac{(l+P)^2}{P\cdot q}-1 \approx \frac{2P\cdot l}{P\cdot q}-1.
\eeq

The hadronic tensor is expressed as a convolution of the reduced hadronic tensors $w_{q,g}^{\mu\nu}$ and the quark and 
gluon FFs $D_{1q,g}(z)$, which describe the processes $q(P_h/z), g(P_h/z) \to h(P_h)$,
\beq
W^{\mu\nu}=\sum_{i=q,g}\int \frac{dz}{z^2}D_{1i}(z)w_i^{\mu\nu}.
\eeq
In the following we will suppress the flavor summation.
The tensor $w^{\mu\nu}$ is represented by the sum of the two diagrams in Fig.~\ref{dia},
\beq
w_{\mu\nu}=\int \frac{d^4k}{(2\pi)^4} {\rm Tr}[M^{(0)}(k)S_{\mu\nu}^{(0)}(k)] + \int \frac{d^4k_1}{(2\pi)^4}\frac{d^4k_2}{(2\pi)^4} {\rm Tr}[M^{(1)}_\sigma (k_1,k_2)S^{(1)\sigma}_{\mu\nu}(k_1,k_2)]. \label{master}
\eeq
The hard matrix elements $S_{\mu\nu}^{(0)}(k)$ and $S_{\mu\nu}^{(1)\sigma}(k_1,k_2)$, with $\sigma$ being 
the polarization index of the attached gluon, can be computed in perturbation theory.
The nonperturbative proton matrix elements $M_{ij}^{(0)}\sim \langle P S_T| \bar{\psi}_j\psi_i|P S_T\rangle$ and
$M_{ij}^{(1)\sigma}\sim \langle P S_T|\bar{\psi}_j gA^\sigma \psi_i|P S_T\rangle$ have indices in Dirac space
($ij$), as well as in color space (omitted for simplicity). Here $S_T^\mu=(0,{\boldsymbol S}_T,0)$ is the spin vector of the transversely polarized proton with the normalization $S_T^2=-1$. 

The collinear factorization approach amounts to expanding the momentum $k^\mu$ in $S^{(0)}$ and 
$k^\mu_{1,2}$ in $S^{(1)}$ around the collinear part proportional to $P^\mu$,
\beq
k^\mu = xP^\mu + k_T^\mu, \qquad k_{1,2}^\mu = x_{1,2}P^\mu + k_{1,2 T}^\mu.
\eeq
After some manipulations, one finds (see Eqs.~(31) and (42) of \cite{Eguchi:2006mc})
 \beq
w_{\mu\nu}&=& \int dx {\rm Tr}\left[M^{(0)}(x) S^{(0)}_{\mu\nu}(x) \right] 
+ \int dx {\rm Tr} \left[ iM^{(0)\alpha}_\partial (x) \frac{\partial S^{(0)}_{\mu\nu}(k)}{\partial k^\alpha_T} \right]_{k=xP}\nn
&+& \int dx_1 dx_2 {\rm Tr} \Biggl[ M^{(1)+}(x_1,x_2)S_{\mu\nu}^{(1)-}(x_1,x_2) \nn
&& \qquad \qquad \qquad  +
 i M_F^{(1)\alpha}(x_1,x_2)\frac{\partial S_{\mu\nu}^{(1)-}(k_1,k_2)}{\partial k_{2 T}^\alpha }  \nn &&
  \qquad \qquad \qquad  + M^{(1)\alpha}(x_1,x_2) \Biggl( P^+ (x_2-x_1)\frac{\partial S_{\mu\nu}^{(1)-}(k_1,k_2)}{\partial k^\alpha_{2T}}+ S^{(1)}_{\alpha\mu\nu}(x_1,x_2)\Biggr) \nn
 && \qquad \qquad \qquad + i M^{(1)\alpha}_{\partial 1}(x_1,x_2) \Biggl( \frac{\partial S_{\mu\nu}^{(1)-}(k_1,k_2)}{\partial k_{1T}^\alpha} +\frac{\partial S_{\mu\nu}^{(1)-}(k_1,k_2)}{\partial k_{2T}^\alpha} \Biggr) \Biggr]_{k_i=x_iP},
\label{ma}
 \eeq
  where $\alpha$ is transverse, $S^{(0)}(x) \equiv S^{(0)}(xP)$, $S^{(1)}_\sigma(x_1,x_2) \equiv S^{(1)}_\sigma(x_1P,x_2P)$, 
  and
    \beq
  M^{(1)\alpha}_\partial (x) =\int \frac{d\lambda}{2\pi}e^{i\lambda x}\langle P S_T|\bar{\psi}(0) 
  \partial^\alpha \psi(\lambda n)|P S_T\rangle,
  \eeq
  \beq
  M^{(1)\alpha}_F(x_1,x_2) =\int \frac{d\lambda}{2\pi}\int \frac{d\zeta}{2\pi} e^{i\lambda x_1 
  + i\zeta(x_2-x_1)}\langle P S_T|\bar{\psi}(0) gF^{\alpha+}(\zeta n)\psi(\lambda n)|P S_T\rangle,
  \eeq
  \beq
  M^{(1)\alpha}(x_1,x_2) =\int \frac{d\lambda}{2\pi}\int \frac{d\zeta}{2\pi} e^{i\lambda x_1 
  + i\zeta(x_2-x_1)}\langle P S_T|\bar{\psi}(0) gA^\alpha(\zeta n)\psi(\lambda n)|P S_T\rangle,
  \eeq
  \beq
  M^{(1)\alpha}_{\partial 1}(x_1,x_2) =\int \frac{d\lambda}{2\pi}\int \frac{d\zeta}{2\pi} 
  e^{i\lambda x_1 + i\zeta(x_2-x_1)}\langle P S_T|\bar{\psi}(0) gA^+(\zeta n) \partial^\alpha_T \psi(\lambda n)|P S_T\rangle,
  \eeq
with $n^\mu=\delta^\mu_-/P^+$. The authors in \cite{Eguchi:2006mc} have focused on the third line of 
Eq.~(\ref{ma}), evaluating the corresponding one-loop hard kernel in perturbation theory and obtaining 
the soft gluon pole (SGP), soft fermion pole (SFP) and hard pole (HP) contributions. They have also 
shown that all the other lines in Eq.~(\ref{ma}) vanish identically for these contributions.

However, all the lines in Eq.~(\ref{ma}) can actually contribute to SSA in more general situations.
It has been pointed out in \cite{Ratcliffe:1985mp,Eguchi:2006mc} that the first line potentially contributes to SSA, 
if one picks up the $g_T$ distribution function
\beq
M^{(0)}(x) = \frac{M_N}{2}\gamma_5\Slash S_T g_T(x) +\cdots,
\eeq
with $M_N$ being the proton mass.
The authors of \cite{Eguchi:2006mc} noted that if $S^{(0)}$ is calculated in the Born (one-loop) approximation, 
the asymmetry trivially vanishes, because there is no phase from the Born diagrams to cancel the $i$ from 
the trace involving $\gamma_5$ (see the discussion in Sec.~\ref{sec:twoloop}). As we shall demonstrate later, 
certain two-loop diagrams for $S^{(0)}$ can generate a phase, which leads to a contribution to SSA proportional 
to $g_T$. When this occurs, the second line of Eq.~(\ref{ma}) provides the ${\cal O}(g)$ piece of the Wilson line 
in the definition of $g_T$. To see this, note the QCD Ward-Takahashi (WT) identity for the contraction of 
a gluon of the momentum $k_2 - k_1$,
\beq
(k_2-k_1)_\sigma S^{(1)\sigma}_{\mu\nu}(k_1,k_2)=S_{\mu\nu}^{(0)}(k_2) -S_{\mu\nu}^{(0)}(k_1), \label{cho}
\eeq
where a color matrix $t^a$ with the color index $a$ is implicit on the right hand side.
The above formula gives
\beq
P^+S^{(1) -}_{\mu\nu}(x_1,x_2) = -\frac{S_{\mu\nu}^{(0)}(x_2)}{x_1-x_2+i\epsilon} 
+ \frac{S_{\mu\nu}^{(0)}(x_1)}{x_1-x_2+i\epsilon},\label{cho1}
\eeq
in the collinear limit. Upon the integration over $x_1$ or $x_2$, the factor $1/(x_1-x_2+i\epsilon)$ becomes the
$\theta$-function that enters the Wilson line integral
\beq
\bar{\psi}(0)\int_0^{\lambda} d\zeta A^+(\zeta n)\psi(\lambda n).
\eeq

Differentiating Eq.~(\ref{cho}) with respect to $k_{1,2}$ and then taking the collinear limit, one finds
\beq
&&  P^+(x_2-x_1)\left. \frac{\partial S_{\mu\nu}^{(1)-}(k_1,k_2)}{\partial k^\alpha_{2T}} \right|_{k_i=x_iP}
+ S^{(1)}_{\mu\nu\alpha}(x_1,x_2) 
=\left. \frac{\partial S^{(0)}_{\mu\nu}(k_2)}{\partial k^\alpha_{2T} }   \right|_{k_i=x_iP}, \nn
&&  \left. P^+(x_2-x_1)\frac{\partial S_{\mu\nu}^{(1)-}(k_1,k_2)}{\partial k^\alpha_{1T}}\right|_{k_i=x_iP} 
- S^{(1)}_{\mu\nu\alpha}(x_1,x_2) 
=-\left. \frac{\partial S_{\mu\nu}^{(0)}(k_1)}{\partial k^\alpha_{1T} }   \right|_{k_i=x_iP}. \label{two}
 \eeq
It means that the fourth line of Eq.~(\ref{ma}) is non-vanishing, and the fifth line does not vanish either, 
as one can see by summing the two relations in Eq.~(\ref{two}).
The crucial difference between the analysis of \cite{Eguchi:2006mc} and ours is whether the right
hand sides of Eq.~(\ref{two}) vanish or not.
Following \cite{Eguchi:2006mc}, the hard kernel $S^{(1)}$ is defined as the sum of `irreducible' diagrams without including the `reducible' diagrams in which the $k_2-k_1$ gluon merges with the incoming or returning quark line.
With this definition, the right hand sides of Eq.~(\ref{two}), accounting for the contributions from 
those reducible diagrams, exist in general. See \cite{Kanazawa:2013uia,Hatta:2013wsa,Xing:2019ovj} 
for related discussions in the context of SSA.
It turns out that, for the SGP, SFP and HP contributions at the Born level considered in \cite{Eguchi:2006mc}, the right hand sides of Eq.~(\ref{two}), and the fourth and fifth lines of Eq.~(\ref{ma}), all vanish.
However, for the set of two-loop diagrams proposed in the next section, the right hand sides Eq.~(\ref{cho}) do not vanish. The fourth and fifth lines of Eq.~(\ref{ma}) do not vanish either, and they must be treated simultaneously for gauge invariance as elaborated below.

Inserting Eq.~(\ref{two}) into Eq.~(\ref{ma}), we observe that various terms organize themselves to form gauge invariant twist-three matrix elements.\footnote{Our notations are the same as in \cite{Eguchi:2006mc}:
$\gamma_5=i\gamma^0\gamma^1\gamma^2\gamma^3$, $\epsilon_{0123}=1$ and $\epsilon^{\alpha P n S_T}\equiv \epsilon^{\alpha\beta\gamma\delta}P_\beta n_\gamma S_{T \delta}$.
%Since we are dealing with the forward matrix elements, we can freely replace the covariant derivative $D_T^\alpha=\partial_T^\alpha-igA_T^\alpha$  with $-\overleftarrow{D}_T^\alpha=-\overleftarrow{\partial}_T^\alpha-igA_T^\alpha$, or more symmetrically, $\overleftrightarrow{D}_T^\alpha = (D_T^\alpha -\overleftarrow{D}_T^\alpha)/2$ in Eq.~(\ref{gd}).
}
Define
\beq
&&\int\frac{d\lambda}{2\pi} \int \frac{d\mu}{2\pi} e^{i\lambda x_1+i\mu(x_2-x_1)} \langle P S_T|\bar{\psi}_j(0)[0,\mu n]D^\alpha_T(\mu n)[\mu n, \lambda n] \psi_i(\lambda n)|P S_T\rangle \nn
&&= \frac{M_N}{4}(\Slash P)_{ij} \epsilon^{\alpha P n S_T} G_D(x_1,x_2) + i\frac{M_N}{4}(\gamma_5\Slash P)_{ij}S^\alpha_T \tilde{G}_D(x_1,x_2), \label{gd}
\eeq
\beq
&&\int\frac{d\lambda}{2\pi} \int \frac{d\mu}{2\pi} e^{i\lambda x_1+i\mu(x_2-x_1)} \langle P S_T|\bar{\psi}_j(0)[0,\mu n] g F^{\alpha \beta}(\mu n)n_\beta [\mu n,\lambda n] \psi_i(\lambda n)|P S_T\rangle \nn
&&= \frac{M_N}{4}(\Slash P)_{ij} \epsilon^{\alpha P n S_T} G_F(x_1,x_2) + i\frac{M_N}{4}(\gamma_5\Slash P)_{ij}S^\alpha_T \tilde{G}_F(x_1,x_2), %\label{gd}
\eeq
where the Wilson line $[\mu n, \lambda n] = {\rm P}\exp\left[ i g \int_\lambda^\mu dt n\cdot A(t n)\right]$ renders 
the matrix elements gauge invariant, and the three-parton PDFs obey the symmetry property,
\beq
G_D(x_1,x_2)=-G_D(x_2,x_1), \qquad \tilde{G}_D(x_1,x_2)=\tilde{G}_D(x_2,x_1), \label{sym} \\
G_F(x_1,x_2)=G_F(x_2,x_1), \qquad \tilde{G}_F(x_1,x_2)=-\tilde{G}_F(x_2,x_1) . \nonumber
\eeq
The second term of the first line and the fourth line in Eq.~(\ref{ma}) combine to give the covariant 
derivative $\bar{\psi}D^\alpha_T\psi$, and the fifth line provides the Wilson line of this operator to 
make it gauge invariant. Equation~(\ref{ma}) then becomes
\beq
w_{\mu\nu}&=&M_N\int dx {\rm Tr} \left[\gamma_5 \Slash S_T \frac{g_T(x)}{2} S^{(0)}_{\mu\nu}(x ) \right]  \label{for} \\
&+& \frac{iM_N}{4}\int dx_1dx_2 {\rm Tr}\left[ \left(\Slash P \epsilon^{\alpha P n S_T}  G_D(x_1,x_2) + i\gamma_5\Slash P S_T^\alpha \tilde{G}_D(x_1,x_2) \right)\left.\frac{\partial S^{(0)}_{\mu\nu}(k)}{\partial k_T^\alpha}\right|_{k=x_2P} \right]\nn
&+& \frac{iM_N}{4}\int dx_1 dx_2 {\rm Tr}\left[\left(\Slash P \epsilon^{\alpha P n S_T}  \frac{G_F(x_1,x_2)}{x_2-x_1} + i\gamma_5\Slash P S_T^\alpha  \frac{\tilde{G}_F(x_1,x_2)}{x_2-x_1} \right) \left(\left. \frac{\partial S^{(0)}_{\mu\nu}(k)}{\partial k_T^\alpha}\right|_{k=x_2P} -S_{\mu\nu\alpha}^{(1)}(x_1,x_2)\right) \right]. \nonumber
\eeq
The above expression can be further simplified by using the identity \cite{Eguchi:2006qz}
\beq
G_D(x_1,x_2)={\cal P}\frac{G_F(x_1,x_2)}{x_1-x_2}, \qquad \tilde{G}_D(x_1,x_2)= \delta(x_1-x_2)\tilde{g}(x_1)+{\cal P}\frac{\tilde{G}_F(x_1,x_2)}{x_1-x_2},  \label{ident}
\eeq
where ${\cal P}$ denotes the principal value prescription. We shall omit ${\cal P}$ below to avoid 
confusion with the momentum $P^\mu$. The second equation can be regarded as the definition of $\tilde{g}(x)$, 
that is in fact related to $g_T(x)$, $G_F$  and $\tilde{G}_F$ through the QCD equation of motion (see 
Eq.~(\ref{la}) below).\footnote{$\tilde{g}(x)$ is related to the first moment $g_{1T}^{\perp (1)}(x)$ of 
the twist-3 TMD $g_{1T}(x,k_T^2)$. We find $\tilde{g}(x) = - 2 g_{1T}^{\perp (1)}(x)$, where the 
definition of $g_{1T}^{\perp (1)}(x)$ from \cite{Kanazawa:2015ajw} has been used.} We thus arrive at
\beq
w_{\mu\nu}&=&\frac{M_N}{2}\int dx g_T(x) {\rm Tr} \left[\gamma_5 \Slash S_T S^{(0)}_{\mu\nu}(x ) \right]  \label{rev} \\
&-& \frac{M_N}{4}\int dx \tilde{g}(x) {\rm Tr}\left[ \gamma_5\Slash P S_T^\alpha  \left.\frac{\partial S^{(0)}_{\mu\nu}(k)}{\partial k_T^\alpha}\right|_{k=xP} \right]\nn
&+& \frac{iM_N}{4}\int dx_1 dx_2 {\rm Tr}\left[\left(\Slash P \epsilon^{\alpha P n S_T}  \frac{G_F(x_1,x_2)}{x_1-x_2} + i\gamma_5\Slash P S_T^\alpha  \frac{\tilde{G}_F(x_1,x_2)}{x_1-x_2} \right)  S_{\mu\nu\alpha}^{(1)}(x_1,x_2) \right], \nonumber
\eeq
which will be the starting point of our two-loop analysis. 
%proof for the gauge invariance.

\subsection{QED gauge invariance}
\label{wt}
Let us show that Eq.~(\ref{rev}) respects the QED WT identity, which is actually nontrivial. 
The WT identity for $S^{(0)}$ is written as
\beq
q^\mu \Slash P S_{\mu\nu}^{(0)}(x) =0, \qquad q^\nu S_{\mu\nu}^{(0)}(x)\Slash P =0 , \label{li}
\eeq
where $\Slash P=P^+\gamma^-$, and $q^\mu$ and $q^\nu$ represent the outgoing and incoming photon momenta, respectively. It is obvious that the first line of Eq.~(\ref{rev}) does not satisfy the WT identity by itself due to the presence of $\gamma_5\Slash S_T$. (For unpolarized distributions, one has the spin projector $\gamma^-$ instead, and the WT identity is trivially satisfied.)
In fact, only the sum of all lines in Eq.~(\ref{rev}) obeys the WT identity.  Similar observations have been made in the literature \cite{Ratcliffe:1985mp,Jaffe:1991ra,Kanazawa:2013uia}.

To verify it, we begin with a slight generalization of Eq.~(\ref{li}),
\beq
q^\mu \Slash k S_{\mu\nu}^{(0)}(k) =0, \qquad q^\nu S_{\mu\nu}^{(0)}(k)  \Slash k=0, \label{on}
\eeq
for an on-shell, but not necessarily collinear momentum $k$.
Differentiating Eq.~(\ref{on}) with respect to $k_T^\alpha$ and then taking the collinear limit, we get
\beq
q^\mu \gamma_{T\alpha} S_{\mu\nu}^{(0)}(x) + \left. q^\mu x\Slash P \frac{\partial S^{(0)}_{\mu\nu}(k)}{\partial k_T^\alpha}\right|_{k=xP}=0, \qquad q^\nu  S_{\mu\nu}^{(0)}(x) \gamma_{T\alpha} + \left. q^\nu \frac{\partial S^{(0)}_{\mu\nu}(k)}{\partial k_T^\alpha}\right|_{k=xP}x\Slash P =0. \label{lim}
\eeq
Furthermore, we need the following identity \cite{Eguchi:2006mc}
\beq
g_T(x) = -\frac{1}{2x} \left( \tilde{g}(x) +  \int dx' \frac{G_F(x,x') +\tilde{G}_F(x,x')}{x-x'} \right), \label{la}
\eeq
where the $\tilde{g}(x)$ part combines with the second line of Eq.~(\ref{rev}) to give the structure
\beq
\sim \int dx \tilde{g}(x) {\rm Tr}\left[ \gamma_5\Slash S_T \frac{S_{\mu\nu}^{(0)}(x)}{x} +\gamma_5 \Slash P S_T^\alpha \left.\frac{\partial S^{(0)}_{\mu\nu}(k)}{\partial k_T^\alpha}\right|_{k=xP} \right] = \int dx \frac{ \tilde{g}(x)}{x} {\rm Tr}\left[ \gamma_5 S_T^\alpha \frac{  \partial (\Slash k S^{(0)}_{\mu\nu}(k))}{\partial k_T^\alpha} \right]_{k=xP}  . \label{gauge1}
\eeq
This combination vanishes when contracted with $q^\mu$ or $q^\nu$, as can be easily checked by using Eq.~(\ref{lim}).

The $G_F$ and $\tilde{G}_F$ terms of Eq.~(\ref{la}) combine with the third line of Eq.~(\ref{rev}) to give the structure
\beq
\sim \int dx_1 dx_2 {\rm Tr}\Biggl[ \left( - \gamma_5\Slash S_T \frac{S^{(0)}_{\mu\nu}(x_1 )}{x_1} 
+ i\Slash P \epsilon^{\alpha PnS_T} S_{\mu\nu\alpha}^{(1)}(x_1,x_2) \right)  \frac{G_F(x_1,x_2)}{x_1-x_2}  \nn
+ \left( - \gamma_5\Slash S_T \frac{S^{(0)}_{\mu\nu}(x_1)}{x_1} 
-\gamma_5\Slash PS^\alpha_T  S_{\mu\nu\alpha}^{(1)}(x_1,x_2) \right) \frac{\tilde{G}_F(x_1,x_2)}{x_1-x_2} \Biggr].
\label{gauge2}
\eeq
Remembering that $S^{(1)}$ does not contain reducible diagrams, we have
\beq
q^\mu  S_{\mu\nu\alpha}^{(1)}(x_1,x_2) =\gamma_\alpha \frac{1}{x_1\Slash P}q^\mu S_{\mu\nu}^{(0)}(x_1), \qquad  q^\nu S^{(1)}_{\mu\nu\alpha}(x_1,x_2)=q^\nu S_{\mu\nu}^{(0)}(x_2)\frac{1}{x_2\Slash P}\gamma_\alpha. \label{bo}
\eeq 
Using the following formulas
\beq
\gamma_5 \Slash S_T = i\gamma^0\gamma^1\gamma^2\gamma^3 \Slash S_T=-i(1-\gamma^+\gamma^-)\gamma_1\gamma_2 (S_{T 1}\gamma_1 + S_{T 2}\gamma_2)  &=& i(1-\gamma^+\gamma^- )\epsilon^{ij}\gamma_i S_{T j}, \label{c1}
\\ &=&  i(\gamma^-\gamma^+ -1 )\epsilon^{ij}\gamma_i S_{T j}, \nonumber
\eeq
\beq
 -i\gamma_\alpha \epsilon^{\alpha -+ \lambda}S_{T \lambda} = i\epsilon^{ij}\gamma_i S_{T j}, \label{c2}
\eeq
together with Eqs.~(\ref{sym}) and (\ref{li}), one can show that both lines of Eq.~(\ref{gauge2}) vanish,
when contracted with $q^\mu$ or $q^\nu$. This completes the proof of the QED WT identity.

\subsection{QCD gauge invariance}

Similarly, the QCD gauge invariance holds only for the sum of all terms in Eq.~(\ref{rev}). Suppose that $S^{(0)}$ 
is evaluated in some gauge which involves a parameter $\xi$ (here we have suppressed the subscripts $\mu$, $\nu$ for 
simplicity). For instance, $\xi$ can be the usual gauge parameter $\lambda$ in the covariant gauge, or a vector 
$n^\alpha$ in the axial gauge $n\cdot A=0$, in which the gluon propagator is proportional to
\begin{eqnarray}
N_{\rm co}^{\alpha\beta}&=&g^{\alpha\beta}-(1-\lambda) \frac{l^\alpha
l^\beta}{l^2}, \nonumber\\
N_{\rm ax}^{\alpha\beta}&=&g^{\alpha\beta}- \frac{l^\alpha n^\beta+l^\beta n^\alpha}{l\cdot n} 
+ n^2\frac{l^\alpha l^\beta}{(l\cdot n)^2},
\label{gp}
\end{eqnarray}
respectively. We will show that Eq.~(\ref{rev}) does not change under the variation of the gauge parameters $\xi$,
concentrating on these two classes of gauges.
To vary the $\lambda$ or $n$ dependence in diagrams at arbitrary orders, we apply the
differential operator $d/d\lambda$ or $d/dn_\delta$ to each of the gluon propagators,
yielding
\begin{eqnarray}
\lambda\frac{d}{d\lambda}N_{\rm co}^{\alpha\beta}&=& \frac{l_\delta}{2l^2}
\left(N_{\rm co}^{\alpha \delta}l^\beta+N_{\rm co}^{\delta \beta}l^\alpha\right),\nonumber\\
\frac{d}{dn_\delta}N_{\rm ax}^{\alpha\beta}&=& -\frac{1}{n\cdot l}
\left(N_{\rm ax}^{\alpha \delta}l^\beta+N_{\rm ax}^{\delta \beta}l^\alpha\right).
\label{digp}
\end{eqnarray}
%where the factors $l_\delta/(2l^2)$ and $-1/n\cdot l$, introduced by the differentiation, are 
%not crucial for the proof of the QCD gauge invariance.

\begin{figure}[!]
\includegraphics[scale=0.5]{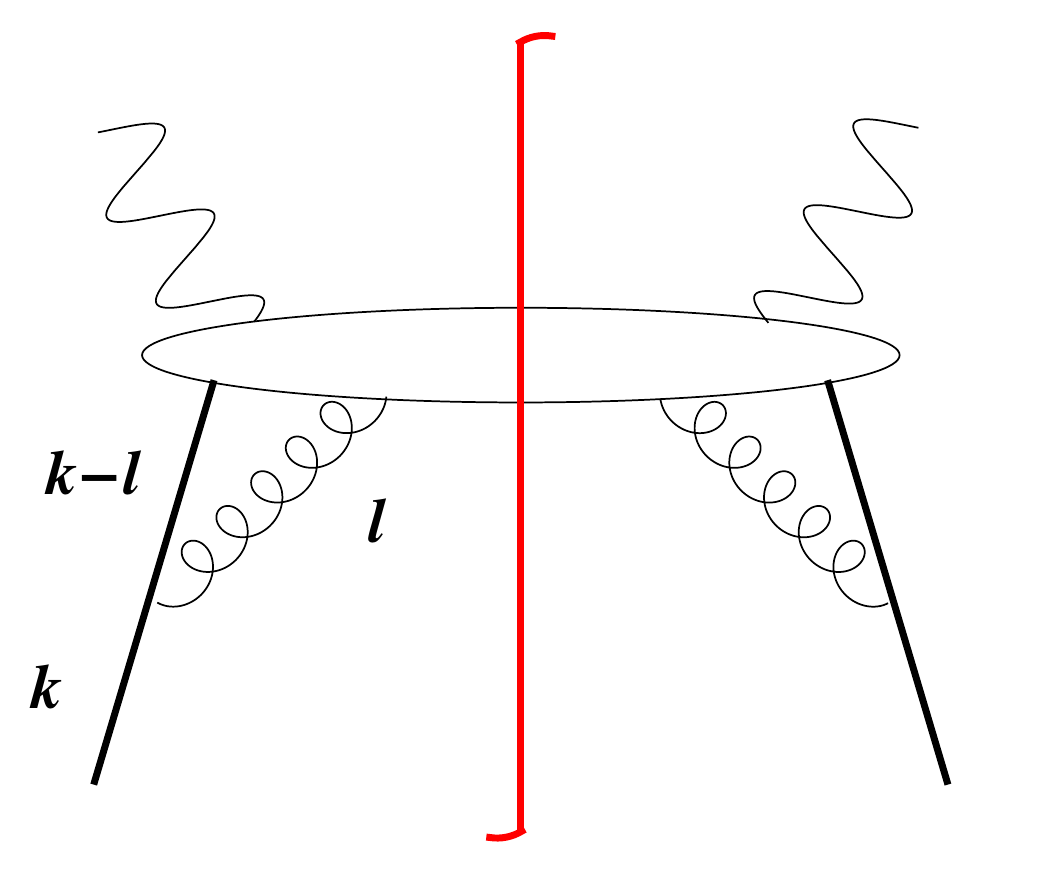}
\caption{The variation of the gauge parameters is represented by the diagram, in which the momentum 
$l^\beta$ attaches to the outermost end of either the incoming or returning quark.}
\label{out}
\end{figure}

Starting with the $\tilde{g}$ terms in Eq.~(\ref{gauge1}), one writes the differentiated 
$S^{(0)}(k,\xi)$ as $\delta S^{(0)}(k,\xi)$. The momentum $l^\alpha$ or $l^\beta$ appearing at one end 
of the differentiated gluon line (\ref{digp}) is contracted with a vertex the gluon attaches to. We select an ordinary
gluon vertex denoted by $\alpha$ (without the contraction with its momentum) in the diagrams, 
and collect vertices which correspond to the attachments of another end denoted by $\beta$. Since all gluons 
are differentiated, the possible attachments of $l^\beta$ form a
complete set of diagrams. Summing all the gluon attachments, one finds that the only uncanceled 
piece comes from the diagram with the momentum attaching to the outermost end of either the 
incoming or returning quark \cite{BS89}, as depicted in Fig.~\ref{out}.  
One thus obtains $\delta S^{(0)}(k,\xi)= \delta S_L^{(0)}(k,\xi)+\delta S_R^{(0)}(k,\xi)$ corresponding 
to these two possibilities.
Clearly they satisfy
\beq
  \delta S_L^{(0)}(k,\xi)\Slash k=0, \qquad \Slash k\delta S_R^{(0)}(k,\xi)=0, \label{cut}
\eeq
which are entirely analogous to Eq.~(\ref{on}). It is then trivial to see that Eq.~(\ref{gauge1}) with $S^{(0)}$ being replaced by $\delta S_{L/R}^{(0)}$ vanishes. Therefore, the $\tilde{g}$ part is gauge independent.

Similarly, one can write the differentiated three-parton amplitude $S^{(1)}$ as
\beq
\delta S_\alpha^{(1)}(x_1,x_2,\xi)=\delta S^{(1)}_{L\alpha}(x_1,x_2,\xi)
+\delta S^{(1)}_{R\alpha}(x_1,x_2,\xi),
\eeq
for which the QCD gauge invariance holds for the sum of the reducible and irreducible diagrams. 
We then have
\beq
  \left(\delta S^{(1)}_{L\alpha}(x_1,x_2,\xi) -\delta S_L^{(0)}(x_2,\xi)  \frac{1}{x_2\Slash P}\gamma_\alpha  \right)\Slash P
  %= \delta S^{(1)}_{L\alpha}(x_1,x_2,\delta\xi)\Slash P  +\frac{\delta S_L^{(0)}(x_2,\delta\xi)}{x_2}  \gamma_\alpha
  =   0, \nn
\Slash P\left(\delta S^{(1)}_{R\alpha}(x_1,x_2,\xi) 
-\gamma_\alpha \frac{1}{x_1\Slash P} \delta S_{R}^{(0)}(x_1,\xi) \right)
%=\Slash P \delta S^{(1)}_{R\alpha}(x_1,x_2,\delta\xi) +\gamma_\alpha \frac{ \delta S_R^{(0)}(x_1,\delta\xi)}{x}
 =0,
\eeq
which are again completely analogous to Eq.~(\ref{bo}). Hence, Eq.~(\ref{gauge2}) with $S^{(0),(1)}$ being 
replaced by $\delta S_{L/R}^{(0),(1)}$ vanishes.
This completes the proof that Eq.~(\ref{rev}) is QCD gauge invariant.

\section{Two-loop contribution to phase}
\label{sec:twoloop}

\begin{figure}[!]
\includegraphics[scale=1]{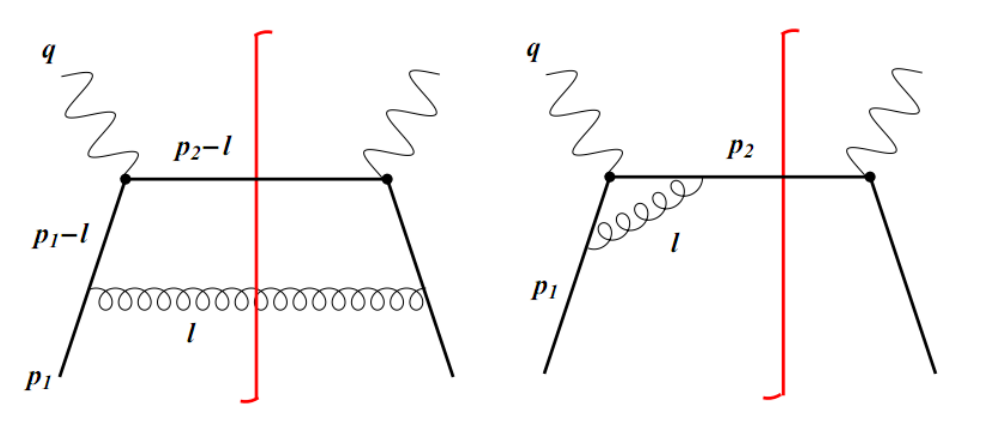}
\caption{Left: A one-loop box diagram. Right: A one-loop vertex-correction diagram.
} \label{1loop}
\end{figure}

In this section, we identify the lowest order two-parton Feynman diagrams that produce nonvanishing contributions to 
Eq.~(\ref{rev}) in the collinear factorization.  It was pointed out \cite{Eguchi:2006mc} that the Born term, 
given by the one-loop box diagram in Fig.~\ref{1loop} (left), does not contribute.
We can easily confirm this result by an explicit calculation as follows.
The incoming quark has the momentum $p_1=xP$ with $1\ge x\ge x_B$, and we write the virtual photon momentum as
$q=p_2-p_1$ with
\beq
p_2^+ = (x-x_B)P^+, \qquad p_2^- = \frac{Q^2}{2x_BP^+}, \qquad p_2^2=\frac{x-x_B}{x_B}Q^2.
\eeq
Figure~\ref{1loop} (left) with the loop gluon momentum $l^\mu=(l^+,l^-,\lT)$ is evaluated as
\beq
%{\color{red}S_{\mu\nu}^{(0),{\rm box}}(x)}&\sim &-g^2C_F
\int\frac{d^4l}{(2\pi)^4}
\frac{\gamma_\alpha(\Slash p_1-\Slash l)\gamma_\mu (\Slash p_2-\Slash l)\gamma_\nu (\Slash p_1-\Slash l) 
\gamma^\alpha}{[(p_1-l)^2+i\epsilon][(p_1-l)^2-i\epsilon]}
\delta((p_2-l)^2)\delta(l^2)
%\nn&\times & {\color{red}(2\pi)^4\delta^{(4)}\left(p_2 - l - \frac{P_h}{z}\right)}
,\label{s0}
\eeq
whose integrand, as contracted with $\gamma_5\Slash S_T $, yields a factor $i$. 
%{\color{red} Here $C_F = (N_c^2 - 1)/(2 N_c) $.} 
In order to make the cross section real, the denominator must provide an imaginary part. However, 
this is clearly not possible, so the one-loop box diagram does not contribute to SSA.

Next, consider the virtual correction
to the photon vertex in Fig.~\ref{1loop} (right),\footnote{In the collinear factorization framework, this diagram does not contribute to SSA trivially, since the final state quark has a vanishing transverse momentum. We nevertheless study the pole structure of this diagram (and other virtual diagrams below) because our discussion can be straightforwardly generalized to the $k_T$ factorization framework, where the incoming quark has a nonzero transverse momentum and the analysis becomes nontrivial.}
\begin{eqnarray}
%{\color{red}{\rm Tr}[\gamma_5 \Slash S_T S_{\mu\nu}^{(0),{\rm vertex}}(x)]}&\sim &-ig^2C_F
\int\frac{d^4l}{(2\pi)^4}
\frac{\gamma_\mu\Slash p_2\gamma^\alpha(\Slash p_2-\Slash l)\gamma_\nu
(\Slash p_1-\Slash l)\gamma_\alpha}
{[(p_2-l)^2+i\epsilon]( l^2+i\epsilon) [(p_1-l)^2+i\epsilon]}\delta(p_2^2)
%(2\pi)^4\delta^{(4)}\left(p_2 - \frac{P_h}{z}\right)
,\label{s1}
\end{eqnarray}
in which the final state quark is on-shell with $p_2^+=0$ ($x=x_B$). The loop
integral over $l$ needs to generate an imaginary
piece in order to get a real contribution. Expressing
$(p_1-l)^2=2(l^+-p_1^+)l^--l_T^2$, $l^2=2l^+l^--l_T^2$, and $(p_2-l)^2=
2l^+(l^--p_2^-)-l_T^2$, we see that $l^+$ must take a value in the range
$(0,p_1^+)$ to get a nonvanishing contribution from the contour integration over $l^-$. 
After picking up the pole $l^-=l_T^2/[2(l^+-p_1^+)]+i\epsilon$, we need one more $i$ from the remaining $l$ or $p_2 - l$ propagator. However, this is impossible due to
$l^2=2p_1\cdot l=2p_1^+l^-<0$ and $(p_2-l)^2=2(p_1\cdot l-p_2\cdot l)=2(p_1^+l^--p_2^-l^+)<0$.
Namely, neither the gluon nor the scattered quark can become on-shell, so this diagram does not contribute.

These observations apply to other one-loop diagrams, and
we conclude that the asymmetry cannot be produced in a parton-level diagram at one loop.

\begin{figure}[!]
\includegraphics[scale=0.5]{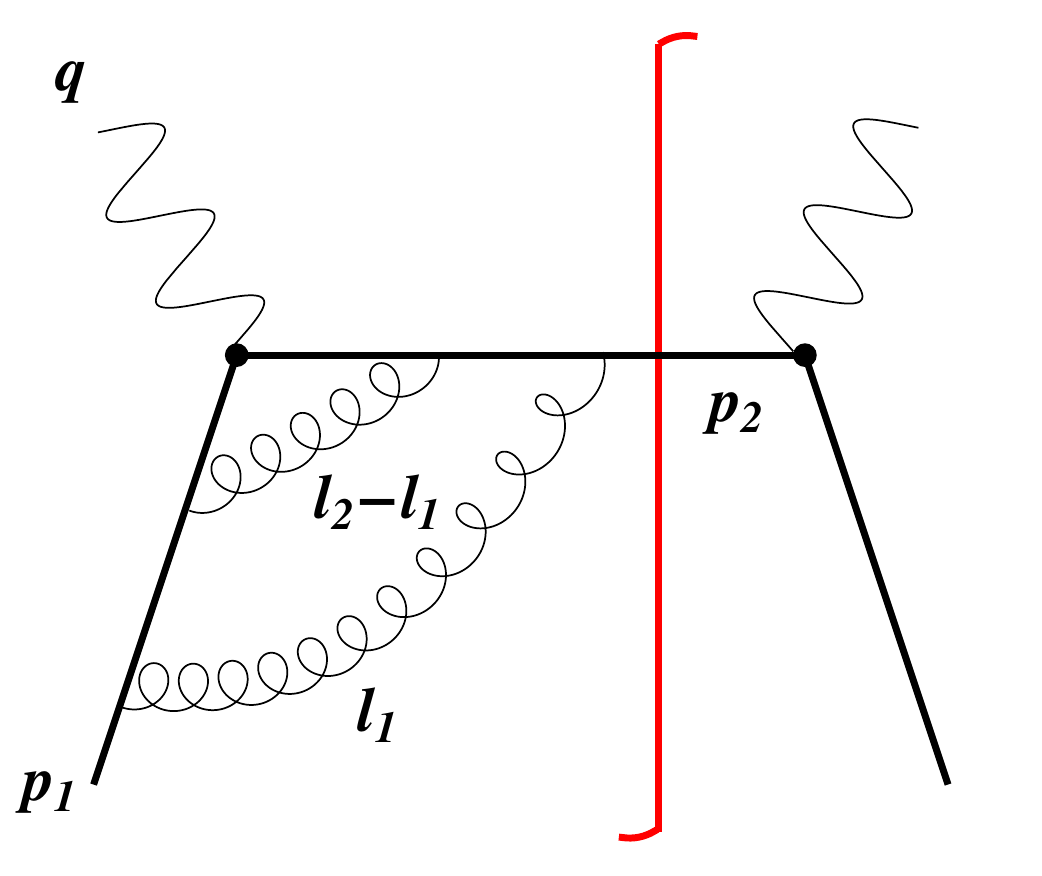}
\caption{A two-loop diagram with two virtual gluons.
} \label{2loop1}
\end{figure}

\subsection{Fig.~\ref{2loop1}: a case with two virtual gluons}

We then move on to two-loop diagrams, starting with
the diagram with two virtual gluons in Fig.~\ref{2loop1} (see also footnote 2).
Let the incoming quark carry the momentum
$p_1-l_1$ after emitting the first gluon, and $p_1-l_2$ after emitting
the second gluon of the momentum $l_2-l_1$.  The scattered quark then carries the momentum 
$p_2-l_2$ before receiving the
second gluon and $p_2-l_1$ before receiving the first gluon. Focus only on the propagator denominators entering the loop integrand for this diagram, and consider the
poles of $l_1^-$ and $l_2^-$ (again, $p_2^+=0$):
\beq
&&\int_{-\infty}^\infty dl_1^- d l_2^- \frac{1}{[(p_1 - l_1)^2 + i\epsilon](l_1^2+ i\epsilon)[(l_1 - l_2)^2+ i\epsilon][(p_1 - l_2)^2+ i\epsilon](l_2^2+ i\epsilon)}\nn
&=& \int_{-\infty}^\infty dl_1^- dl_2^- \frac{1}{[2(l_1^+-p_1^+)l_1^- -l_{1T}^2+i\epsilon](2l_1^+l_1^- - l_{1T}^2 +i\epsilon)[2(l_1^+-l_2^+)(l_1^--l_2^-)-(\loT - \ltT)^2+i\epsilon]} \nn
&&\times \frac{1}{[2(l_2^+-p_1^+)l_2^- - l_{2T}^2 +i\epsilon](2l_2^+l_2^- -l_{2T}^2 +i\epsilon)}.
\eeq
It is easy to see that as long as one of the components $l_1^+$ and $l_2^+$ is greater than
$p_1^+$, the integration over either $l_1^-$ or $l_2^-$
vanishes because the integration contour is not pinched.
For example, if $l_1^+,l_2^+>p_1^+$, all the poles are in the lower-half plane except the one from the propagator $(l_1-l_2)^2$. The coefficient $l_1^+-l_2^+$
is either positive or negative, and then the integration over either $l_1^-$ or
$l_2^-$ vanishes. The same conclusion is drawn, as one of the components
$l_1^+$ and $l_2^+$ is negative.
We thus need to examine only the ranges $0<l_{1,2}^+<p_{1}^+$. 

We first investigate the case with $l_1^+<l_2^+$, for which we pick up the
pole $l_2^-=l_{2T}^2/[2(l_2^+-p_1^+)]+i\epsilon$ from the incoming quark
propagator. As for  the pole of $l_1^-$, we pick up either $l_1^-=l_{1T}^2/[2(l_1^+-p_1^+)]+i\epsilon$
from the incoming quark propagator or
\beq
l_1^-=\frac{(\loT-\ltT)^2}{2(l_1^+-l_2^+)}+l_2^-+i\epsilon
=\frac{(\loT-\ltT)^2}{2(l_1^+-l_2^+)}+\frac{l_{2T}^2}{2(l_2^+-p_1^+)}+i\epsilon, \label{sec}
\eeq
from the gluon propagator with the momentum $l_2 - l_1$. 
The first pole of $l_1^-$ does not lead to any on-shell internal particles, which all
have negative invariant masses as $l_1^+<l_2^+$. Indeed,
the invariant masses of the scattered quark are given by $(p_2-l_1)^2=2p_1^+l_1^--2p_2^-l_1^+<0$
and $(p_2-l_2)^2=2p_1^+l_2^--2p_2^-l_2^+<0$. The two gluons have the invariant
masses $l_1^2=2p_1^+l_1^-<0$ and
\begin{eqnarray}
(l_2-l_1)^2 = l_2^2 - 2 l_1\cdot l_2 + 2p_1\cdot l_1 =  - \frac{l_2^+ - p_1^+}{l_1^+ - p_1^+}
\left(\loT - \frac{l_1^+ - p_1^+}{l_2^+ - p_1^+}\ltT\right)^2 < 0.
\end{eqnarray}
For the second pole of $l_1^-$ in Eq.~(\ref{sec}), we just need to check the incoming quark of the momentum
$p_1-l_1$:
\begin{eqnarray}
(p_1-l_1)^2 = -2p_1 l_1 + 2 l_1 \cdot l_2 - l_2^2 = \frac{l_2^+ - p_1^+}{l_1^+ - l_2^+}
\left(\loT - \frac{l_1^+ - p_1^+}{l_2^+ - p_1^+}\ltT\right)^2 > 0.
\end{eqnarray}
That is, this incoming quark does not go on shell.

We then analyze the case with $l_1^+>l_2^+$, for which we pick up the
pole $l_1^-=l_{1T}^2/[2(l_1^+-p_1^+)]+i\epsilon$ from the incoming quark
propagator. As to the pole of $l_2^-$, we pick up either $l_2^-=l_{2T}^2/[2(l_2^+-p_1^+)]+i\epsilon$
from the incoming quark or
\beq
l_2^-=\frac{(\ltT-\loT)^2}{2(l_2^+-l_1^+)}+l_1^-+i\epsilon
=\frac{(\ltT-\loT)^2}{2(l_2^+-l_1^+)}+\frac{l_{1T}^2}{2(l_1^+-p_1^+)}+i\epsilon,
\eeq
from the second gluon propagator. The discussion is completely analogous to the $l_1^+<l_2^+$ case: one can show 
that none of the remaining propagators can go on-shell, so they cannot produce a phase. We conclude that 
Fig.~\ref{2loop1} does not contribute to SSA.

\begin{figure}[!]
\includegraphics[scale=0.5]{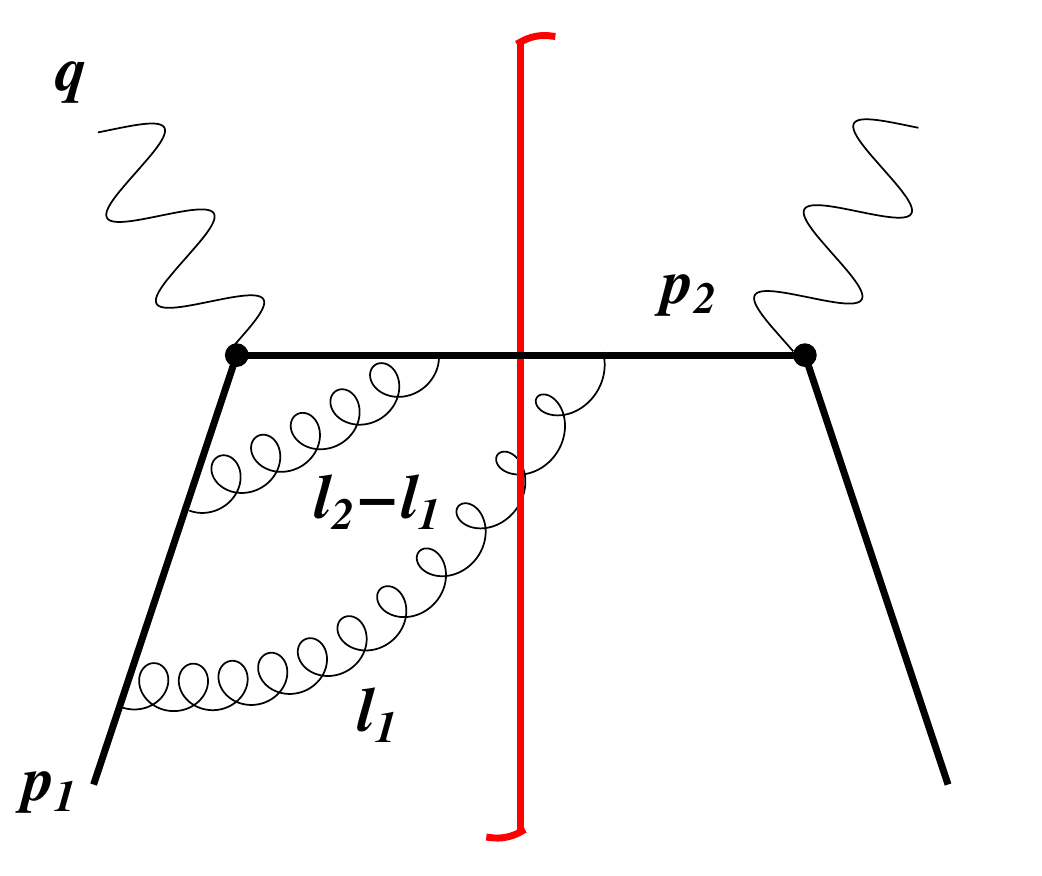}
\hspace{1.0cm}
\includegraphics[scale=0.5]{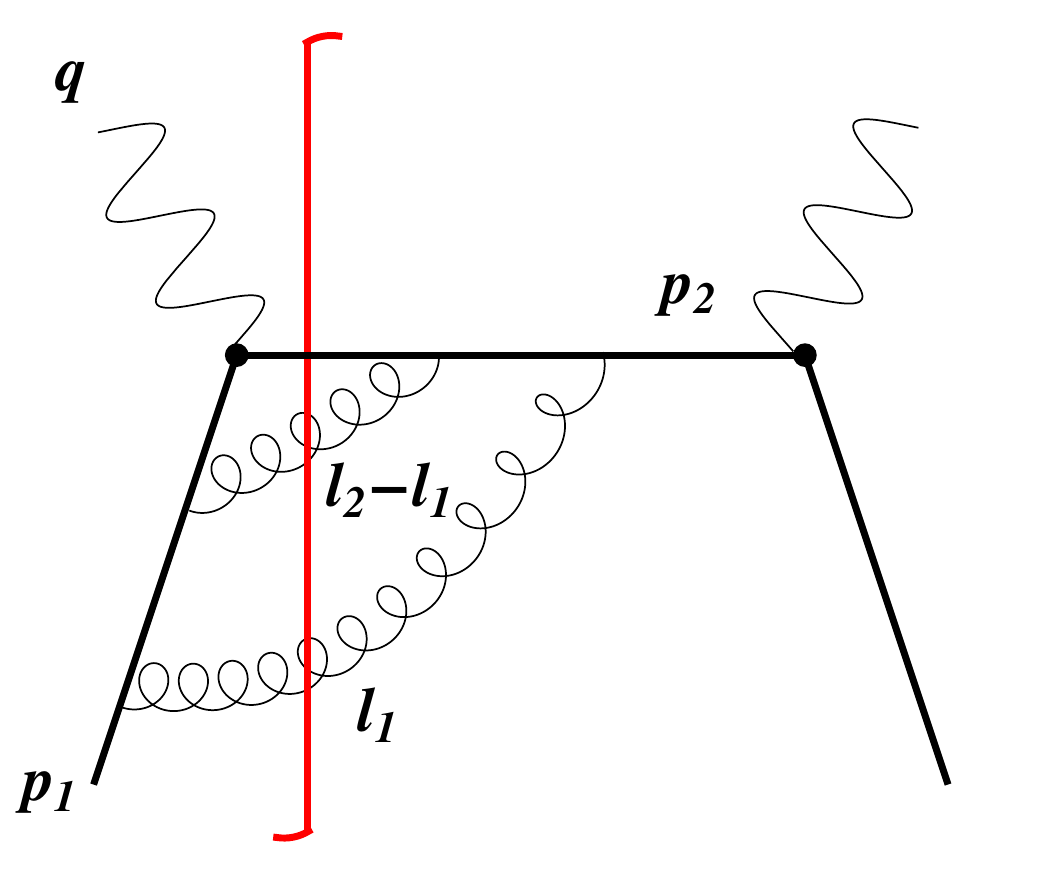}
\caption{Left: A two-loop diagram with one virtual gluon and one real gluon. Right: A two-loop diagram with two real gluons. The phases cancel between these two diagrams (see the text).
} \label{2loop2}
\end{figure}

\subsection{Fig.~\ref{2loop2}: a case of real-virtual cancellation}

When one gluon is real and another is virtual, there is a chance to get an on-shell parton.
Consider the diagram in Fig.~\ref{2loop2} (left), which has the same assignment of momenta as in Fig.~\ref{2loop1} 
but with a different cut. Because $p_2^2=2p_1\cdot q(1-x_B)\geq 0$,
the scattered quark with the invariant mass $(p_2-l_2)^2=p_2^2-2p_2\cdot l_2$
may go on-shell and generate a phase. Hence, this diagram deserves a careful investigation.

The on-shell condition $l_1^2=0$ leads to $l_1^-=l_{1T}^2/(2l_1^+)$. The on-shell condition
$(p_2-l_1)^2=p_2^2-2p_2\cdot l_1=0$ then yields two solutions
\begin{eqnarray}
l_1^+= \frac{p_2^+}{2}(1 \pm \Delta_1) \equiv l^+_{1(\pm)}, \qquad 
\Delta_1 \equiv \sqrt{1-\frac{4l_{1T}^2}{p_2^2}}, \qquad l^-_{1(\pm)}\equiv \frac{l_{1T}^2}{2l^+_{1(\pm)}},
\label{2m0}
\end{eqnarray}
for which the incoming quark is off-shell by $(p_1-l_1)^2=-2p_1\cdot l_1=-2p_1^+l_1^-<0$.
We then come to the contour integration over $l_2^-$,
\beq
&&\int_{-\infty}^\infty dl_2^- \frac{1}{[(l_2 - l_1)^2 + i\epsilon][(p_1 - l_2)^2 + i\epsilon][(p_2 - l_2)^2 + i\epsilon]}\nn
&=& \int dl_2^- \frac{1}{[2(l_2^+-l_1^+)(l_2^- - l_1^-)-(\loT - \ltT)^2+i\epsilon][2(l_2^+-p_1^+)l_2^- - l_{2T}^2 +i\epsilon][2(l_2^+-p_2^+)(l_2^--p_2^-) - l_{2T}^2 +i\epsilon]},
\eeq
which vanishes for $l_2^+>p_1^+$ as before. For $p_2^+<l_2^+<p_1^+$, we pick up the pole 
$l_2^-=l_{2T}^2/[2(l_2^+-p_1^+)]+i\epsilon$,
that renders both the scattered quark and the virtual gluon off-shell with negative
invariant masses. For $l_1^+<l_2^+<p_2^+$, we pick up the pole
\begin{eqnarray}
l_2^-=\frac{(\ltT-\loT)^2}{2(l_2^+-l_1^+)}+\frac{l_{1T}^2}{2l_1^+}-i\epsilon,
\label{2m}
\end{eqnarray}
which makes the incoming quark of the momentum $p_1-l_2$ off-shell with a negative mass.
The invariant mass of the scattered quark
\beq
(p_2-l_2)^2 = p_2^2 - 2p_2 \cdot l_2 + 2 l_1 \cdot l_2 - l_1^2
%&=&2(l_2^+-p_2^+)\left[\frac{|{\bf l}_{2T}-{\bf l}_{1T}|^2}
%{2(l_2^+-l_1^+)}+\frac{l_{1T}^2}{2l_1^+}-p_2^-\right]-l_{2T}^2\nonumber\\
%&=&
= \frac{l_1^+-p_2^+}{l_2^+-l_1^+}\left(\ltT-\frac{l_2^+-p_2^+}{l_1^+-p_2^+}\loT\right)^2
-\frac{(l_2^+-p_2^+)p_2^+}{(l_1^+-p_2^+)l_1^+}l_{1T}^2-2(l_2^+-p_2^+)p_2^-,\label{2m1}
\eeq
approaches plus infinity as $l_2^+\to l_1^+$ from above, and
$-l_{2T}^2$ as $l_2^+\to p_2^+$. That is, we have an on-shell internal particle,
and an imaginary piece. However, this phase will
be cancelled by a phase from the diagram with two real gluons, which we turn to next.

The diagram in Fig.~\ref{2loop2} (right) with the final state cut on the outgoing quark of the
momentum $p_2-l_2$ and the gluons of the momenta $l_1$ and $l_2-l_1$ is closely related to the previously considered 
diagram. The on-shell conditions $(l_2-l_1)^2=0$ and $(p_2-l_2)^2=0$ are
equivalent to Eq.~(\ref{2m}) and the vanishing of Eq.~(\ref{2m1}), respectively. To get an imaginary
piece, the outgoing quark of the momentum $p_2-l_1$ should go on shell, which then
leads to the condition in Eq.~(\ref{2m0}). Therefore, this diagram can give rise to a phase from the same set of on-shell propagators  as in the diagram of Fig.~\ref{2loop2} (left). It has been known that the
contributions from on-shell partons cancel between virtual and
real corrections. A simple explanation for this cancellation is as follows:
for $l_1^+<l_2^+$, the contour integration over the pole of the gluon propagator
with the momentum $l_2-l_1$ in the 
diagram on the left of Fig.~\ref{2loop2} gives the metric tensor $-g^{\mu\nu}$ of the same sign
as the real gluon in the diagram on the right. The other pieces in the loop integrands also contain the same
sign between the two diagrams. The only difference comes from
the sign of the scattered quark propagators: for the diagram on the left, the quark
propagator with the momentum $p_2-l_2$ is proportional to
\begin{eqnarray}
\frac{1}{(p_2-l_2)^2+i\epsilon}=\frac{1}{[(p_2-l_1)-(l_2-l_1)]^2+i\epsilon}=
\frac{1}{-2(p_2-l_1)\cdot(l_2-l_1)+i\epsilon}.
\end{eqnarray}
For the diagram on the right, the quark propagator with the momentum $p_2-l_1$ is proportional to
\begin{eqnarray}
\frac{1}{(p_2-l_1)^2-i\epsilon}=\frac{1}{2(p_2-l_1)\cdot(l_2-l_1)-i\epsilon},
\end{eqnarray}
where we have used the on-shell conditions $(p_2-l_2)^2=(l_1-l_2)^2=0$.
Hence, the diagram on the right generates the same imaginary piece
as the diagram on the left but with an opposite sign. Summing these diagrams,
the imaginary pieces cancel. The same observation applies to
other diagrams, where the real gluon of the momentum $l_1$ attaches to the incoming quark
on the right hand side of the final state cut. In summary, the sum of
the diagrams with two real gluons and those with one real gluon and one
virtual gluon does not contribute to  SSA.

\begin{figure}[!]
\includegraphics[scale=0.5]{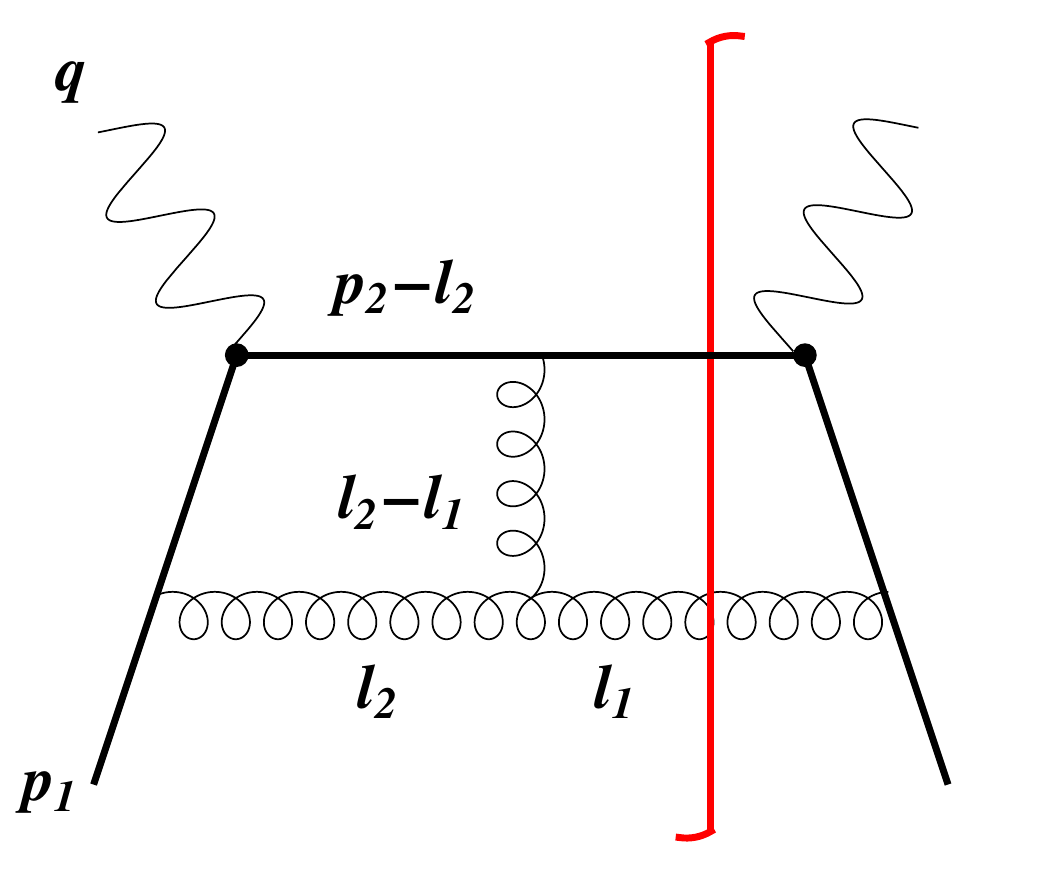}
\caption{The box diagram.
} \label{2loop4}
\end{figure}

\subsection{Fig.~\ref{2loop4}: a two-loop box diagram}

\begin{figure}[!]
\includegraphics[scale=1]{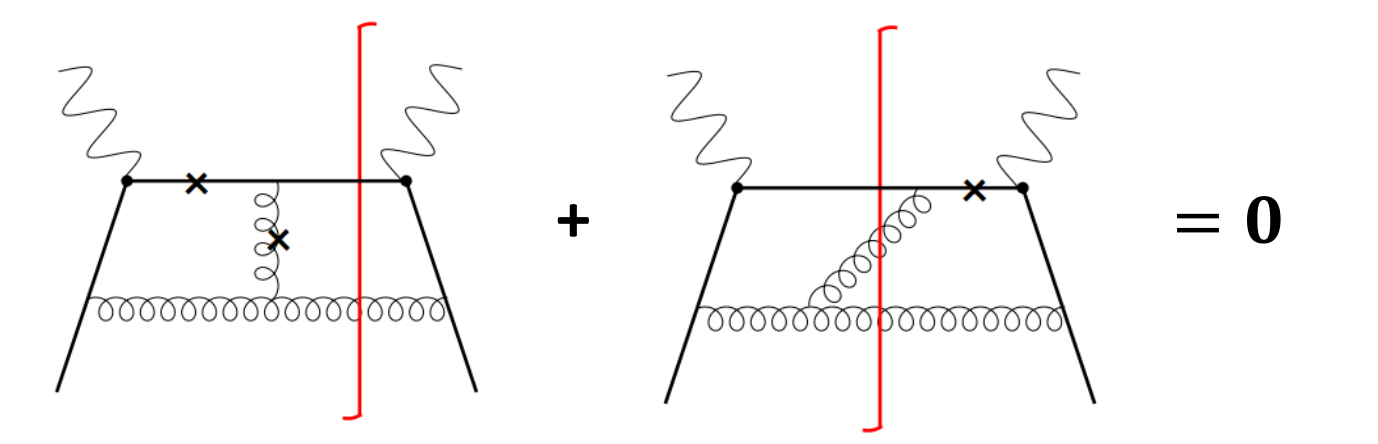}
\caption{Cancellation of particular pole contributions, similar to the one between the two diagrams in Fig.~\ref{2loop2}. Crosses denote on-shell propagators, which give rise to a phase.
} \label{cancel}
\end{figure}

Next we discuss a two-loop box diagram in Fig.~\ref{2loop4},
where two final state partons form a
time-like invariant mass with rescattering between them via
a  virtual gluon with momentum $l_2-l_1$ \cite{Brodsky:2002cx}. The plus and minus components of $l_1$ are
fixed by the final state on-shell conditions as in Eq.~(\ref{2m0}). The contour integration over $l_2^-$ has the structure
\beq
&&\int_{-\infty}^\infty dl_2^- \frac{1}{[(p_1 - l_2)^2 + i\epsilon][(p_2 - l_2)^2 + i\epsilon][(l_2 - l_1)^2 + i\epsilon](l_2^2  + i\epsilon)}\nn
&=& \int_{-\infty}^\infty dl_2^- \frac{1}{[2(l_2^+-p_1^+)l_2^- -l_{2T}^2 +i\epsilon][2(l_2^+-p_2^+)l_2^--l_{2T}^2 +i\epsilon]}\nn
&& \times\frac{1}{[2(l_2^+-l_1^+)(l_2^- - l_1^-)-(\ltT - \loT)^2+i\epsilon](2l_2^+l_2^- - l_{2T}^2+i\epsilon)}.
\eeq
For $p_2^+<l_2^+<p_1^+$, we have $l_2^+ - l_1^+ > 0$, as $l_1^+ < p_2^+$ implied by 
Eq.~(\ref{2m0}). In this case the pole $l_2^-=l_{2T}^2/[2(l_2^+-p_1^+)]+i\epsilon$
renders the outgoing quark $p_2-l_2$ and the two virtual gluons all off-shell with negative
invariant masses. In the range $l_1^+<l_2^+<p_2^+$, we pick up the contributions from two poles, Eq.~(\ref{2m})
and $l_2^-=l_{2T}^2/(2l_2^+)-i\epsilon$. The former leads to an imaginary piece
from the outgoing quark propagator $p_2-l_2$ shown in Eq~(\ref{2m1}). For this pole,
the incoming quark is off-shell by a negative
invariant mass, and the virtual gluon of the momentum $l_2$ is off-shell by
\begin{eqnarray}
l_2^2 = 2l_1\cdot l_2 - l_1^2 =
\frac{l_1^+}{l_2^+-l_1^+}\left(\ltT-\frac{l_2^+}{l_1^+}\loT\right)^2>0.
\end{eqnarray}
Following the same  reasoning as before, the above imaginary piece
will be canceled by the same type of diagram with the final state
cut on the outgoing quark of the momentum $p_2-l_2$ and the gluon of the momentum
$l_2-l_1$ (see Fig.~\ref{cancel}).

\begin{figure}[!]
\includegraphics[scale=1]{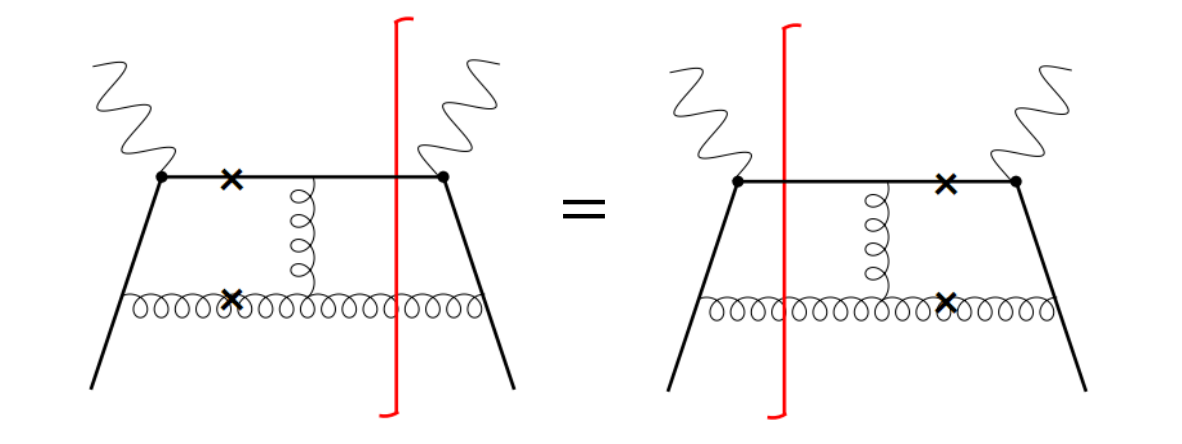}
\caption{Non-cancellation of poles between diagrams with different final state cuts.
} \label{add}
\end{figure}

\begin{figure}[!]
\includegraphics[scale=0.5]{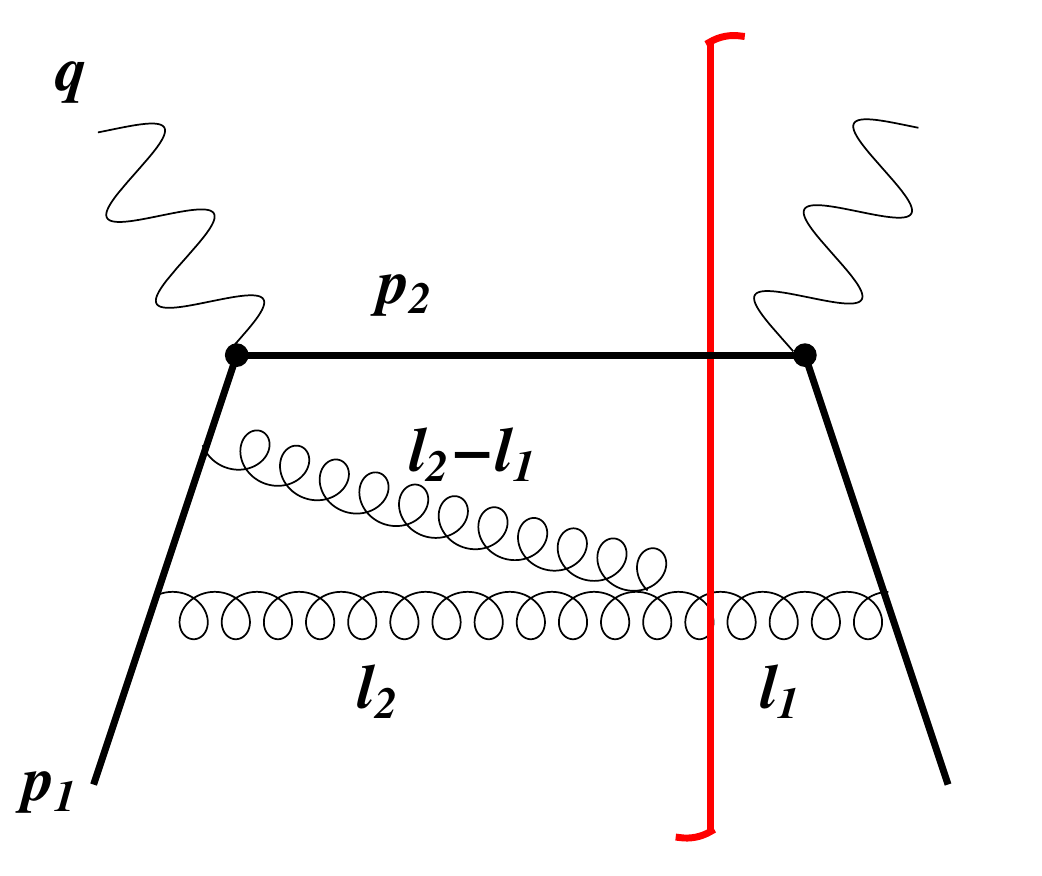}
\caption{A diagram which does not contain a phase.
} \label{2loop5}
\end{figure}

The contribution from the latter pole of $l_2^-$ can be combined with the same pole
in the range $0<l_2^+<l_1^+$, which makes the incoming quark off-shell by a negative
invariant mass, and the virtual gluon of the momentum $l_2-l_1$ off-shell by
\begin{eqnarray}
(l_2 - l_1)^2 = -2l_1\cdot l_2 = -\frac{l_2^+}{l_1^+}\left(\loT-\frac{l_1^+}{l_2^+}\ltT\right)^2 < 0.
\end{eqnarray}
For this pole, the outgoing quark of the momentum $p_2-l_2$ also generates an imaginary
piece, since the on-shell condition $(p_2 - l_2)^2$
can be satisfied. The two solutions are given by
\begin{eqnarray}
l_2^+= \frac{p_2^+}{2}(1 \pm \Delta_2) \equiv l^+_{2(\pm)}, 
\qquad \Delta_2 \equiv \sqrt{1-\frac{4l_{2T}^2}{p_2^2}}, \qquad l^-_{2(\pm)}\equiv \frac{l_{2T}^2}{2l^+_{2(\pm)}},
\label{2p}
\end{eqnarray}
meaning that the imaginary piece persists for arbitrary $l_{1T}^2,l_{2T}^2 <p_2^2/4$.
Note that this contribution is not canceled by the same type
of diagram with the final state cut on the outgoing quark of the momentum
$p_2-l_2$ and the gluon of the momentum $l_2$ (see Fig.~\ref{add}). This diagram is just the
complex conjugate of the considered diagram, and thus gives the identical
contribution. The observation is that we need two final state partons to form
a time-like invariant mass, which rescatter with each other via
exchange of a virtual gluon. The diagram with the virtual gluon of the momentum $l_2-l_1$ attaching to
the incoming quark and the real gluon, displayed in Fig.~\ref{2loop5}, does not contribute an 
imaginary piece: as the first emitted gluon
$l_2$ becomes on-shell, the second emitted gluon $l_2-l_1$ is off-shell and the loop integral 
does not produce a phase.

\begin{figure}[!]
\includegraphics[scale=0.4]{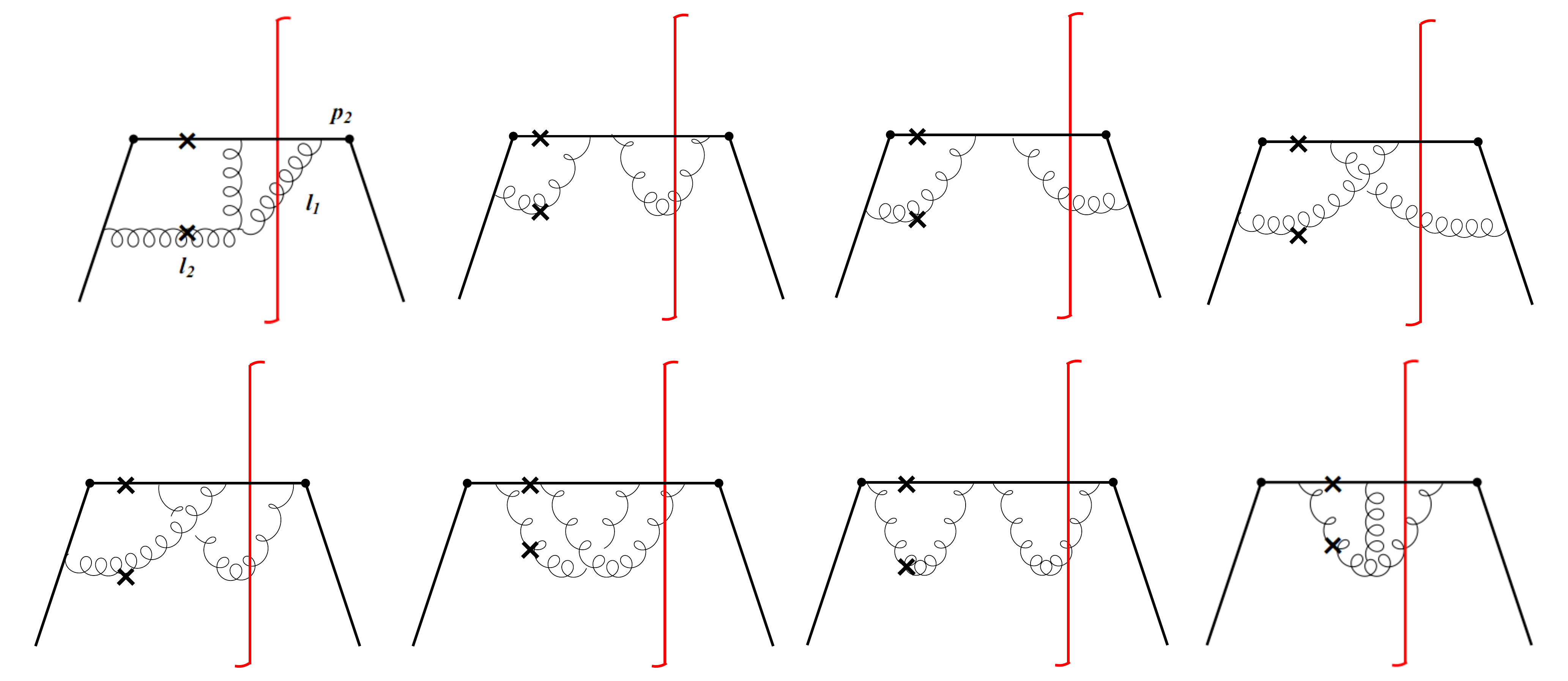}
\caption{Two-loop diagrams which have the same pole structure as the box diagrams. The diagrams obtained by the left-right mirror reflection of the first, second and fifth diagrams are omitted. Virtual photon lines are also omitted for simplicity.
} \label{2loop6}
\end{figure}

\begin{figure}[!]
\includegraphics[scale=0.5]{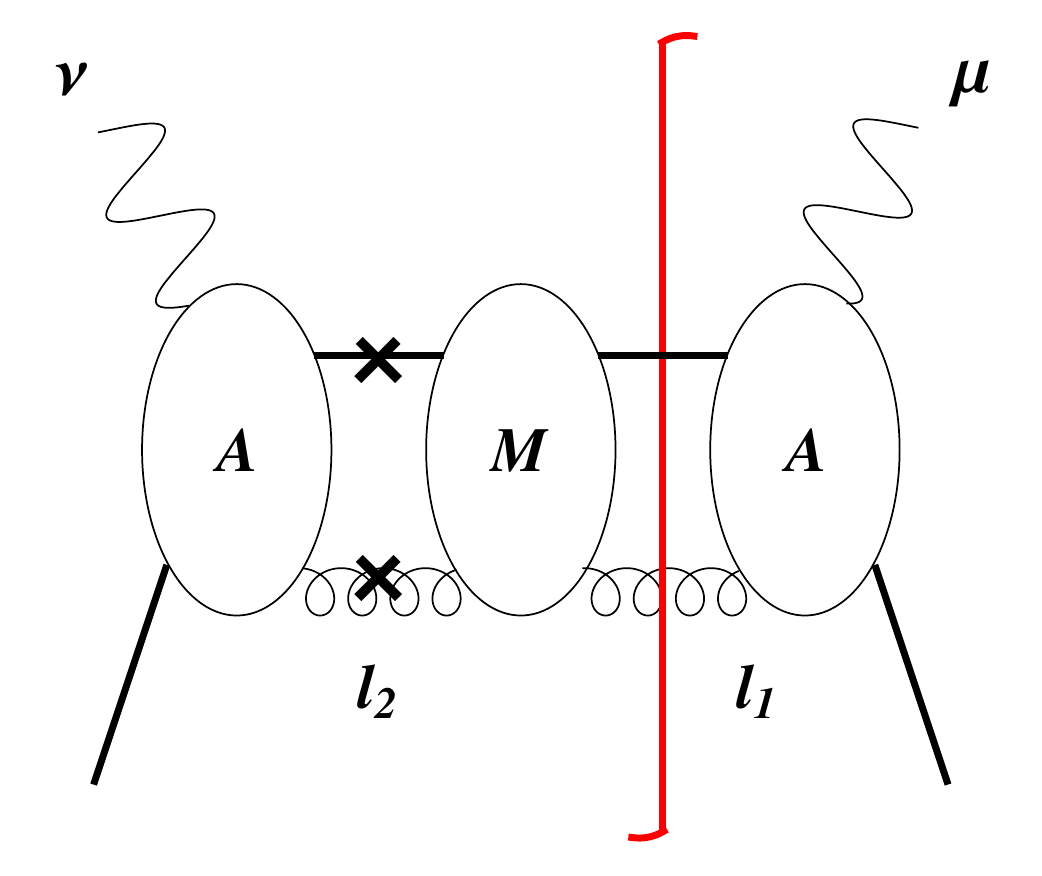}
\caption{A graphical representation of the first term in Eq.~(\ref{hs}).
} \label{blob}
\end{figure}

There exists a class of diagrams, as displayed in Fig.~\ref{2loop6}, which have exactly the same set of on-shell 
propagators as in Fig.~\ref{add}, and are equally important. 
The first and eighth diagrams can be directly obtained from the box diagram by changing the photon vertices. They guarantee that the imaginary piece alone respects the QED WT identity.\footnote{One might think that Fig.~\ref{2loop5} is also related to the box diagram via the WT identity. However, as we have argued, this diagram does not contain a phase, so the WT identity is satisfied without it.} The other diagrams, such as the third and fourth diagrams in the first row, are obtained from the box diagram by changing the attachments of the $l_2$ gluon. They are thus crucial for the QCD gauge invariance. The sum of all these diagrams is written as the following compact formula, as depicted in Fig.~\ref{blob},
\beq
S^{(0)\mu \nu}(x)& =& \frac{g^4}{N_c}(2\pi)\delta\left(\left(p_2 - \frac{P_h}{z}\right)^2\right)\int \frac{d^4 l_2}{(2\pi)^4} (2\pi)\delta(l_2^2)(2\pi) \delta((p_2 - l_2)^2)\nn
&&\times \left\{ i A^{\alpha\mu}(l_1) \Delta_\alpha^{\ \alpha'}M_{\alpha'\beta}(l_1,l_2)A^{\nu\beta}(l_2)-iA^{\alpha\mu}(l_2) M_{\alpha\beta}(l_2,l_1)\Delta^\beta_{\ \beta'}A^{\nu\beta'}(l_1)\right\},  \label{hs}
\eeq
with the number of colors $N_c$, and $l_1 = p_2 - P_h/z$ being determined by the overall momentum conservation.
$\Delta^\alpha_{\ \alpha'}$ is the projector onto the physical polarization states for the final state gluon $l_1$,
 \beq
 \Delta^{\alpha\alpha'}= \sum_{i=1,2} \epsilon_i^\alpha \epsilon^{*\alpha'}_i= -g^{\alpha\alpha'} + \frac{l_1^\alpha \bar{l}_1^{\alpha'}+l_1^{\alpha'}\bar{l}_1^\alpha}{l_1\cdot \bar{l}_1},
 \eeq
with $l=(l^0,\vec{l})$ and  $\bar{l}=(l^0,-\vec{l})$. As long as we sum over all the terms in Eq.~(\ref{rev}) to ensure the gauge invariance, we may replace $\Delta^{\alpha\alpha'}$ by $-g^{\alpha\alpha'}$.
The other factors in Eq.~(\ref{hs}) are defined as
\beq
M_{\alpha\beta}(l_1,l_2)&=&(\Slash p_2-\Slash l_1) t^a \left[-i f^{abc}t^c\frac{V_{\alpha\beta\rho}\gamma^\rho}{(l_1-l_2)^2} + t^at^b\frac{\gamma_\alpha\Slash p_2\gamma_\beta}{p_2^2} + t^bt^a\gamma_\beta
\frac{\Slash p_2 -\Slash l_1-\Slash l_2}{(p_2-l_1-l_2)^2}\gamma_\alpha \right]t^b (\Slash p_2 -\Slash l_2) , \label{mm}
\eeq
\beq
V_{\alpha\beta\rho}&=& g_{\alpha\beta}(l_2+l_1)_\rho + g_{\alpha\rho}(l_2-2l_1)_\beta + g_{\rho\beta}(l_1-2l_2)_\alpha,
\label{vert}
\eeq
and
\beq
A^{\alpha\mu}(l_{1})=
\gamma^\alpha \frac{(\Slash p_1-\Slash l_{1})}{(p_1-l_{1})^2}\gamma^\mu +\gamma^\mu\frac{\Slash p_2}{p_2^2}\gamma^\alpha,
\eeq
\beq
A^{\nu\beta}(l_{2})=
\gamma^\nu \frac{(\Slash p_1-\Slash l_{2})}{(p_1-l_{2})^2}\gamma^\beta +\gamma^\beta\frac{\Slash p_2}{p_2^2}\gamma^\nu.
\eeq
The two terms in Eq.~(\ref{hs}) correspond to the two possible insertions of the final state cut (cf. Fig.~\ref{add}). Taking the hermitian conjugate of the second term, one confirms that Eq.~(\ref{hs}) is symmetric in the indices $\mu,\nu$.

It should be noted that, in the end, the final set of diagrams are identical to those considered in \cite{Ma:2008cj}. We have however provided a more complete analysis of diagrams, including the discussion of gauge invariance and various kinematic configurations. In particular, we have identified the roots in Eqs.~(\ref{2m0}) and (\ref{2p}) which are essential for the factorization of our new contribution to be highlighted in the next section.  

\subsection{Collinear splitting diagrams}

\begin{figure}[!]
\includegraphics[scale=0.5]{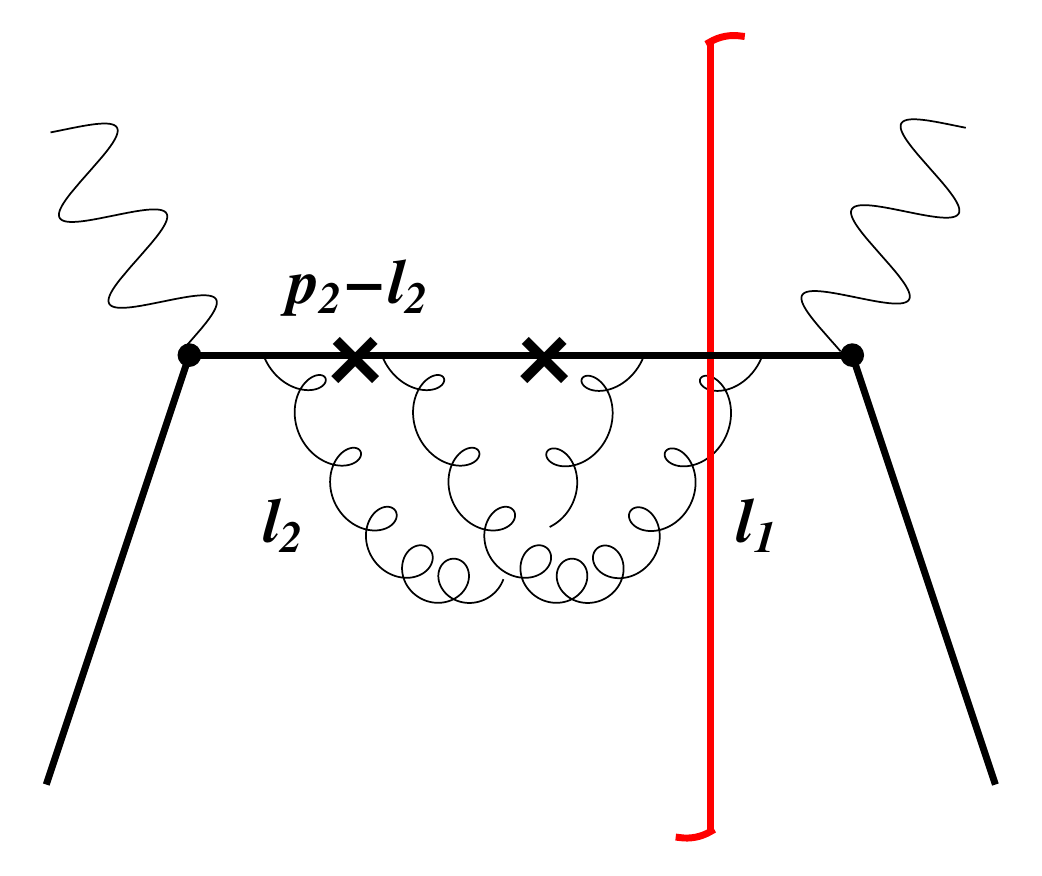}
\caption{A  diagram with an on-shell parton splitting.
} \label{another}
\end{figure}

There exists another class of two-loop diagrams, which contains an imaginary part and is characterized by the collinear splitting of an on-shell parton. An example is shown in Fig.~\ref{another}, where the quark with the momentum $p_2-l_2$ is on-shell, and splits into two on-shell partons, a quark with the momentum $p_2-l_1-l_2$ and a gluon with the momentum $l_1$. This configuration is kinematically possible only if the three partons are all collimated to each other, and thus gets phase space suppression. Indeed, a simple analysis indicates that the imaginary part arises, only if $\ltT$ is opposite in direction relative to $\loT$ and $l_{2T}^2<l_{1T}^2$. It means that this diagram is suppressed by $l_{1T}^2/p_2^2 \sim P_{hT}^2/Q^2$, namely, a higher twist effect. We therefore neglect these diagrams.

\section{Factorization}

Equation~(\ref{hs}) derived in the previous section cannot be immediately inserted into Eq.~(\ref{rev}), 
because it involves collinear divergences from different kinematic regions. In this section we examine 
the structure of these divergences and discuss their treatments in the collinear and $k_T$ factorizations. 
The on-shell conditions for the final state partons and the integration over the light-cone components 
of $l_2$ lead to a summation over the following combinations of roots, see Eqs.~(\ref{2m0}) and (\ref{2p}),
\beq
(l_1^+,l_2^+)=(l_{1(+)}^+,l_{2(+)}^+), \quad (l^+_{1(+)},l^+_{2(-)}), \quad (l^+_{1(-)},l^+_{2(+)}), \quad (l^+_{1(-)},l^+_{2(-)}).
\label{compo}
\eeq
For each choice, the corresponding minus components are fixed by $l^-_{1}=l_{1T}^2/(2l^+_{1})$ and 
$l^-_{2}=l_{2T}^2/(2l^+_{2})$.
We introduce the shorthand notations $(++),(+-),(-+),(--)$ to represent the above four choices. 

\subsection{Collinear factorization}

Since the momentum $l_1$ has been set to $l_1=p_2 - P_h/z$ in the collinear factorization, we  
investigate only the infrared divergence from the integration over $\ltT$.
First consider the $(++)$ and $(-+)$ cases, for which the radiative $l_2$ gluon is collimated to
the initial proton in the collinear region
\begin{eqnarray}
l_{2}^+\sim O(p_2^+)\gg l_{2T} \gg l_{2}^-.\label{h++}
\end{eqnarray}
The incoming quark of the momentum $p_1-l_2$ is nearly on-shell, and the associated $l_{2T}$ integral is 
logarithmically divergent like
\beq
\int \frac{d^2 \ltT}{(p_1-l_2)^2} = \int \frac{d^2 \ltT}{-2p_1^+l_{2+}^-} 
\sim \int \frac{d^2 \ltT}{l_{2T}^2}. \label{coll}
\eeq
The $l_2-l_1$ propagators for the $(\pm +)$ combinations are written as 
\beq
\frac{1}{(l_1-l_2)^2} = \frac{-1}{p_2^2(\Delta_1\mp \Delta_2)^2/4+(\loT-\ltT)^2}.\label{12}
\eeq
There is an apparent divergence at $\loT \to \ltT$ in the $(++)$ case, but it is innocuous because the numerator 
of Eq.~(\ref{hs}) vanishes as $l_1=l_2$.    
The last term of Eq.~(\ref{mm}) is given by
\beq
\frac{1}{(p_2-l_1-l_2)^2} = \frac{-1}{p_2^2(\Delta_1\pm \Delta_2)^2/4 +(\loT+\ltT)^2},
\label{appa}
\eeq
for which the $(-+)$ combination appears problematic in the limit $\loT\to -\ltT$. Inspecting the 
numerator, we find that all components of $p^\mu_2-l^\mu_1-l^\mu_2$ 
go to zero simultaneously as $\loT\to -\ltT$, so this limit is in fact infrared finite.

To determine the nature of the collinear configuration in the $(\pm +)$ combinations, 
look at the potentially dangerous term in Eq.~(\ref{hs}),
\beq
M_{\alpha\beta}A^{\nu\beta} \sim M_{\alpha\beta} \gamma^\nu 
\frac{\Slash p_1-\Slash l_2}{l_{2T}^2}\gamma^\beta. \label{beta}
\eeq
In the small $l_{2T}$ limit, $l_2^\mu$ has only the plus component.
We then immediately see that the $\beta=-$ component in Eq.~(\ref{beta}) vanishes owing to 
$(\Slash p_1-\Slash l_2)\gamma^- \sim (\gamma^-)^2=0$.  As for the component $\beta=+$, we find 
from Eq.~(\ref{mm})
\beq
M_{\alpha +} \propto M_{\alpha \beta}l^\beta_2 \propto l_{1\alpha},
\eeq
which is a consequence of the QCD WT identity. When the longitudinal momentum $l_{1\alpha}$ goes 
into the final state cut, this contribution also vanishes. Therefore, we only need to worry about 
the case, where $\beta$ in Eq.~(\ref{beta}) is transverse.

\begin{figure}[!]
\includegraphics[scale=0.4]{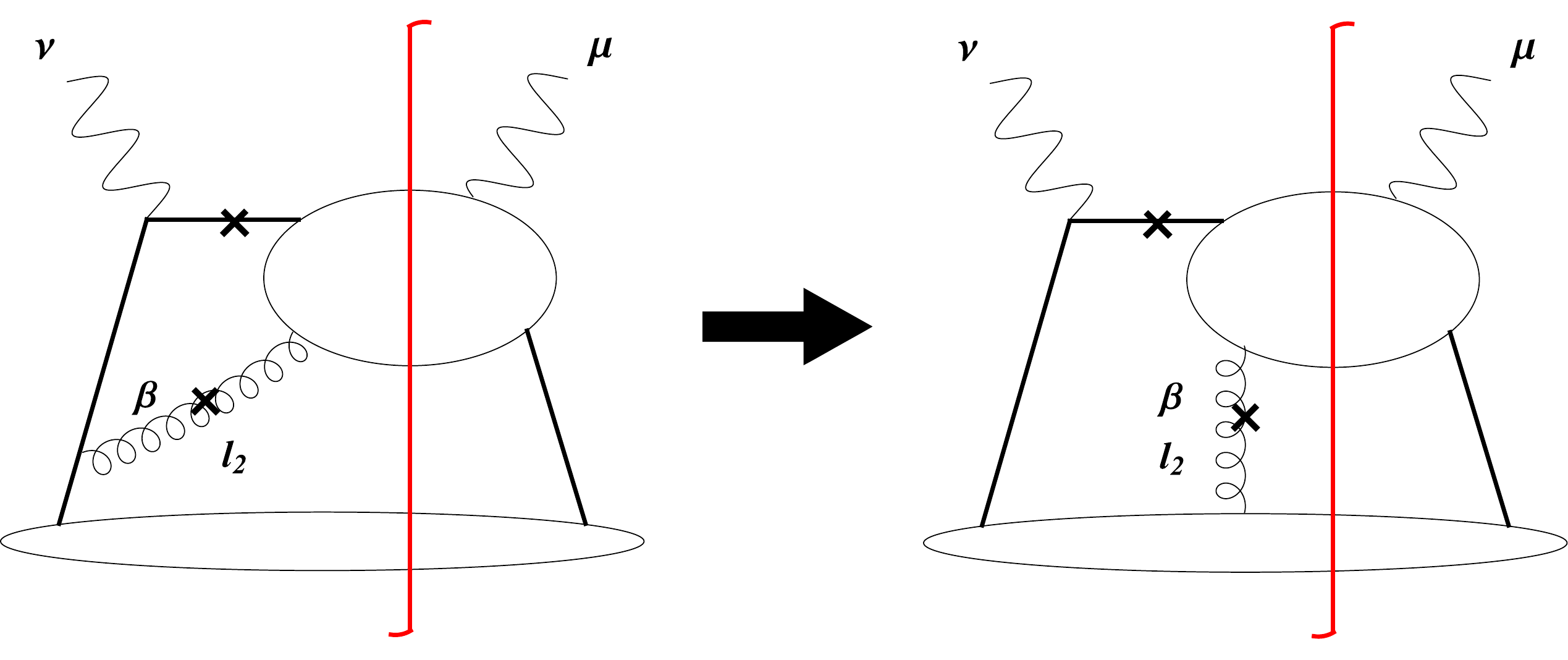}
\caption{The HP contribution from the transversely polarized $l_2$ gluon.
} \label{hp}
\end{figure}

For transverse $\beta$, the singularity does survive. We argue that this can be absorbed into the HP 
contribution to SSA known in the literature.    
Indeed, since the collinear gluon with the momentum 
$l_2$ is transversely polarized and travels a long distance, we may deform Fig.~\ref{blob} into 
Fig.~\ref{hp}, which is identical to Fig.~2 of \cite{Eguchi:2006mc}. As demonstrated in 
\cite{Eguchi:2006mc}, this corresponds to the HP contribution associated with the three-parton ETQS function 
$G_F(x_1,x_2)$, where the value of $x_1$ is set to the Bjorken variable $x_B$: 
label the longitudinal momentum of the 
incoming quark by $p_1^+-l_2^+ = xP^+ -l_2^+ =x_1P^+$ and the gluon momentum by $l_2^+=(x_2-x_1)P^+$. 
The on-shell condition $l_2^+ \approx p_2^+ =(x-x_B)P^+$ then yields $x_1=x_B$. 
In practice, to absorb the collinear divergence into the ETQS function, we insert the projector 
$(\gamma^+)(\gamma^-)$ from the Fierz identity
\begin{eqnarray}
I_{ij}I_{lk}&=& \frac{1}{4}I_{ik}I_{lj}
+ \frac{1}{4}(\gamma^\alpha)_{ik}(\gamma_\alpha)_{lj}
+ \frac{1}{8}(\gamma_5\sigma^{\alpha\beta})_{ik}
(\sigma_{\alpha\beta}\gamma_5)_{lj}
+ \frac{1}{4}(\gamma_5\gamma^\alpha)_{ik}(\gamma_\alpha\gamma_5)_{lj}
+ \frac{1}{4}(\gamma_5)_{ik}(\gamma_5)_{lj},
\label{fierz0}
\end{eqnarray}
into the quark lines with the momenta $p_1-l_2$ and $p_1$ on the left and right hand sides of the cut, respectively.
The matrix $\gamma^+$ then appears as the spin projector in the definition of
the ETQS function, and $\gamma^-$ is contracted to the corresponding one-loop three-parton hard kernel. This factorization has been explicitly demonstrated for a quark target model in \cite{Ma:2008cj}.  We therefore subtract this divergence, as well as the finite part by scheme choice, from Eq.~(\ref{hs}) as a known mechanism. 

%That is, the $(\pm +)$ combinations contribute to the last term of Eq.~(\ref{rev})
%in the collinear factorization framework.

Next we turn to the $(+ -)$ and $(--)$ combinations. The $l_1-l_2$ and the $p_2 - l_1 - l_2$ propagator 
denominators have the forms as in Eqs.~(\ref{12}) and (\ref{appa}), respectively, which are infrared finite in 
the limits $\loT \to \ltT$ and $\loT \to -\ltT$ as explained above. Besides, the radiative $l_2$ gluon satisfies 
the hierarchy
\begin{eqnarray}
l_{2}^-\sim O(p_2^-)\gg l_{2T} \gg l_{2}^+,\label{h--}
\end{eqnarray}
for these combinations, such that there is no infrared singularity in the $p_1-l_2$ propagator. 
Hence, the corresponding phase cannot 
be absorbed into nonperturbative distribution functions. It thus represents a new perturbative 
origin of SSA purely attributed to a hard kernel, and this is the central observation of our work.
In the collinear factorization framework, one can insert the projector $(\gamma^+)(\gamma^-)$ 
between the upper two blobs in Fig.~\ref{dia} and $(\gamma_5\gamma^i)(\gamma_i\gamma_5)$ between 
the lower two blobs. The former leads to the standard collinear twist-two FF $D_1$ and 
the latter leads to the $g_T$ distribution function. We then arrive at a factorization formula
\begin{eqnarray}
d\sigma^{(2)}&=&g_T^{(0)}\otimes H_{\gamma_5\gamma^y,\gamma^+}^{(2)}\otimes D_1^{(0)}, \label{our}
\end{eqnarray}
where the proton spin has been assumed to be along the $y$ direction. The superscript denotes the 
order to which various factors are evaluated.  This is the explicit structure we advocated in Eq.~(\ref{rev}). 

There is, however, another possibility. One can insert the projector 
$(\gamma_5\sigma^{i+})(\sigma_{i+}\gamma_5)$ between the lower two blobs in 
Fig.~\ref{dia} and the identity matrix $(I)(I)$ between the upper two blobs. The former gives the
twist-two transversity distribution function $h_1$, and the latter gives the collinear twist-three 
FF $E$ \cite{Bacchetta:2006tn}. We thus acquire an additional contribution
 \beq
d\sigma^{(2)}= h_1^{(0)}\otimes H_{\gamma_5\sigma^{y-},I}^{(2)}\otimes E^{(0)}. \label{sup}
 \eeq
The FF $E$ dropped out in the one-loop calculation of SSA in SIDIS 
\cite{Kanazawa:2013uia}, where it was denoted as $\hat{e}_1$, and also in $pp$ collisions  
\cite{Metz:2012ct}. It first shows up at two-loops, and is naturally suppressed by a factor 
$\alpha_s$ compared to the one-loop contributions to SIDIS in \cite{Kanazawa:2013uia}.
We point out that an analysis of the complete set of collinear FFs 
is considerably more complicated at twist-three level.  

Of course, Eq.~(\ref{our}) is also parametrically suppressed by a factor $\alpha_s$ compared 
with the known one-loop contributions from the ETQS (or Sivers) distributions \cite{Chen:2017lvx}. The reason we 
nevertheless consider them worthwhile to study is because the $g_T$ distribution function has 
the Wandzura-Wilczek part \cite{Wandzura:1977qf} related to the twist-two polarized quark 
distribution function $\Delta q(x)$. This can be seen from Eq.~(\ref{ident}) together with another 
identity (see Eq.~(45) of \cite{Eguchi:2006qz})
\beq
\tilde{g}(x) =-x\int_x^1 dx_1 \left[ \frac{2\Delta q(x_1)}{x_1} + \frac{1}{x_1^2}\int_{-1}^1 dx_2 
\left( \frac{G_F(x_1,x_2)}{x_1-x_2} +(3x_1-x_2)\frac{\widetilde{G}_F(x_1,x_2)}{(x_1-x_2)^2}\right)\right] .
\eeq
As suggested in \cite{Kanazawa:2014dca}, the genuine twist-three distributions $G_F$ and 
$\widetilde{G}_F$, which are poorly constrained from the experimental data at present, may be 
numerically small. On the other hand, the polarized quark distributions, being purely twist-two 
quantities and well constrained by data, give a finite contribution to the proton spin. 
Hence, the apparent suppression by $\alpha_s$ could be numerically compensated in practice. 
This possibility will be explored in future works \cite{prep}.

The above argument suggests that only the $(\pm -)$ roots is kept in the matrix elements $S^{(0)}$ in Eq.~(\ref{rev}). Remarkably, however, we can include also the $(\pm +)$ roots in
this formula by inserting the Fierz identity into the $p_1$ quark lines,
instead of the $p_1-l_2$ and $p_1$ quark lines, as we have done in the $(\pm -)$ case. 
%In general, it is incorrect to do so, since the `hard kernel' $S_0$ then contains infrared divergences from the colliner $p_1-l_2$ quark. 
%Remarkably, however, we can in fact include also the $(\pm +)$ roots in this formula. As we have seen, $S^{(0)}$ etc. contain infrared divergence from the collinear $p_1-l_2$ quark in the $(\pm +)$ case, and it is in general incorrect to insert the Fierz identity into the $p_1$ quark lines, instead of the $p_1-l_2$ and $p_1$ quark lines, as we have done in the $(\pm -)$ case. 
%If the Fierz identity was inserted 
%into the $p_1$ quark lines, instead of the $p_1-l_2$ and $p_1$ quark lines stated before, all the three terms
%in Eq.~(\ref{rev}) receive the contribution from the $(\pm +)$ combinations. The resultant matrix elements 
%$S^{(0)}$, $\partial S^{(0)}/\partial k_T^\alpha$ and $S_{\alpha}^{(1)}$ then contain the infrared divergence 
%in Eq.~(\ref{coll}) from the $p_1-l_2$ quark. For example, $S_{\alpha}^{(1)}$ collects the three-parton 
%diagrams, in which the valence gluon attaches any internal line in the two-loop $S^{(0)}$.
It will be demonstrated that these %spurious  
divergences due to the alternative Fierz insertion
cancel between the first two terms in Eq.~(\ref{rev}). Substituting Eq.~(\ref{la}) into Eq.~(\ref{rev}), we 
obtain the structures in Eqs.~(\ref{gauge1}) and (\ref{gauge2}). We then notice that 
\beq
S^{(0)}(k) \Slash k \sim M_{\alpha\beta} \gamma^\nu \frac{\Slash k - \Slash l_2}{(k-l_2)^2} \gamma^\beta \Slash k,
\label{IR1}
\eeq
is free of the collinear divergence for an on-shell but not necessarily collinear momentum $k$: in the collinear 
region where $k$ and $l_2$ are parallel, the numerator can be expressed as
\beq
(\Slash k -\Slash l_2)\gamma^\beta \Slash k = 2(k^\beta-l_2^\beta)\Slash k 
-\gamma^\beta (\Slash k-\Slash l_2) \Slash k.\label{IR2}
\eeq 
This gives a vanishing contribution when $k^\beta \propto l_2^\beta$, because of $k^2=0$ and 
$M_{\alpha\beta} l_2^\beta=0$. The differentiation of Eq.~(\ref{IR1}) with respect to 
$S_T^\alpha \partial/\partial k^\alpha$ then immediately leads to the cancellation of the divergences 
in the Wandzura-Wilczek part of $g_T$ in Eq.~(\ref{gauge1}).

\begin{figure}[!]
\includegraphics[scale=0.4]{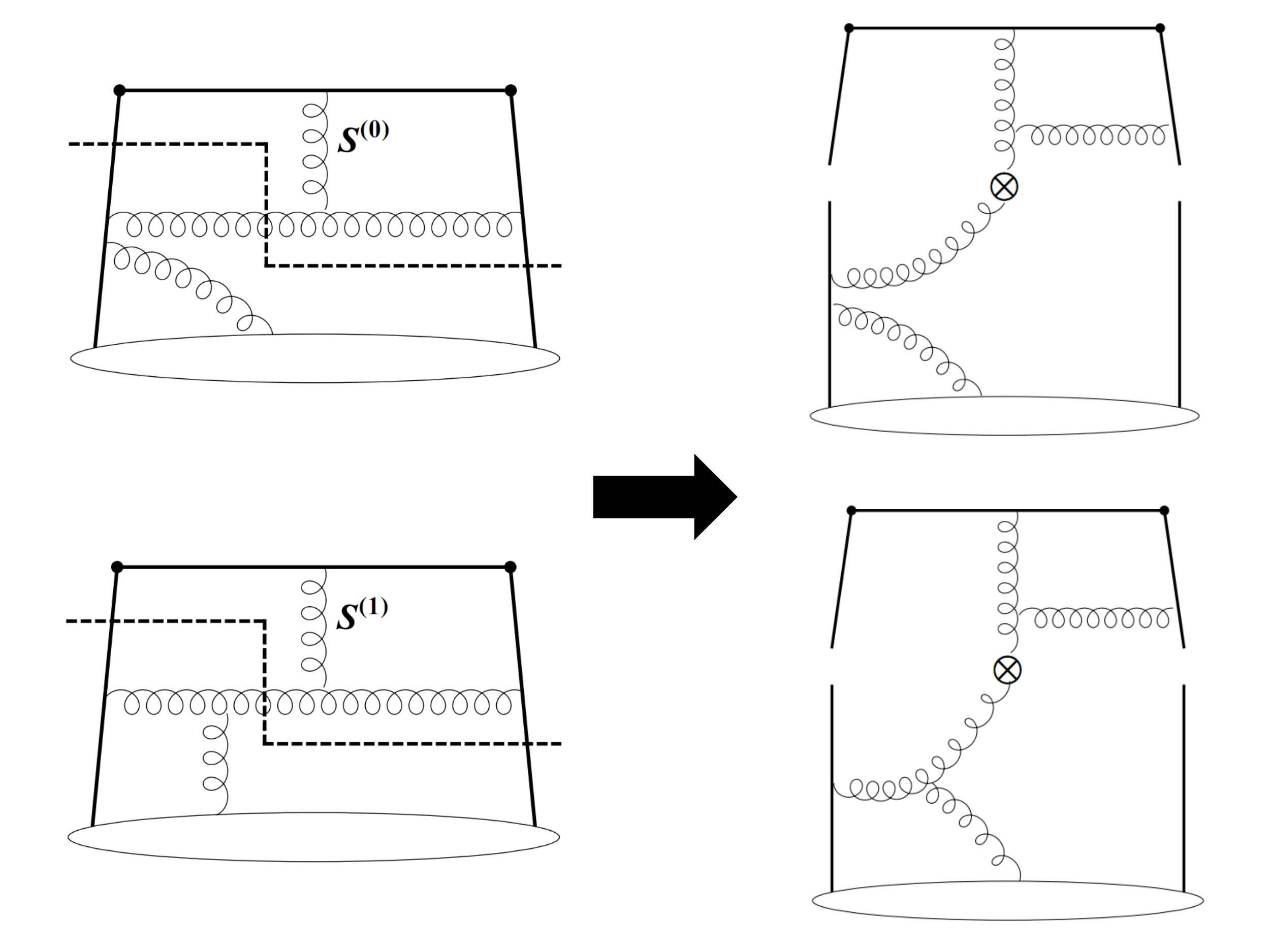}
\caption{A sample of diagrams involved in the factorization of the divergent contribution in Eq.~(\ref{gauge2}), based on the two-loop box diagram. The dashed curve represents the proper insertion of the Fierz identity. 
} \label{fact}
\end{figure}

%After the above cancellation of divergences in the Wandzura-Wilczek part of $g_T$, we are left with the 
%genuine three-parton piece in Eq.~(\ref{gauge2}). 
Including the $(\pm +)$ roots into $S^{(0)}$ and 
$S^{(1)}$, which collects the diagrams with an additional valence gluon attaching to an internal line
of $S^{(0)}$, we find that the resulting collinear divergences do not cancel in Eq.~(\ref{gauge2}). 
We argue that they should be absorbed into the renormalization of the $G_F$ and $\tilde{G}_F$ distributions 
associated with the one-loop HP contribution to SSA. Indeed,  Eq.~(\ref{gauge2}) can be redrawn as in 
Fig.~\ref{fact} by inserting the Fierz identity at a different location. To achieve it, the projectors 
for the $S^{(0)}$ terms have been made the same as for the $S^{(1)}$ terms in the first and second lines of 
Eq.~(\ref{gauge2}) via the replacements $\gamma_5\Slash S_T/x_1=\Slash P\gamma_5\Slash S_T/(x_1\Slash P)
=i\Slash P\gamma_\alpha \epsilon^{\alpha -+ S_T}/(x_1\Slash P)$ and 
$\Slash S_T/x_1=-\Slash PS_T^\alpha\gamma_\alpha/(x_1\Slash P)$, respectively.
In the above expressions $\gamma_\alpha$ corresponds to the vertex located at the outermost end of the 
incoming quark in Fig.~\ref{fact}, and $1/(x_1\Slash P)$ represents the quark propagator following this 
vertex. The lower parts of the diagrams on the right are then identified as the one-loop diagrams to
renormalize the $G_F$ and $\tilde{G}_F$ distributions (see Fig.~7 of Ref.~\cite{Kang:2008ey}). In 
principle, one is able to rederive the evolution equations of $G_F$ and $\tilde{G}_F$ this way. 
We leave it to a future work.

%resulting from the alternative Fierz insertion. Start with the QCD WT identity in Eq.~(\ref{cho}),
%\begin{eqnarray}
%(k_2^\alpha-x_1P^\alpha) S^{(1)}_{\alpha}(x_1P,k_2)\Slash k_2
%+S^{(0)}(x_1P)\Slash k_2-S^{(0)}(k_2)\Slash k_2=0, \label{bo1}
%\end{eqnarray}
%for the contraction of the valence gluon momentum $k_2-x_1P$, where the returning quark 
%momentum $k_2$ is on-shell but not necessarily collinear. Similar to Eq.~(\ref{IR1}), the last
%term in the above identity is infrared finite, so is the sum of the first two terms. We differentiate 
%the first two terms with respect to $S_T^\alpha \partial/\partial k_2^\alpha$, 
%and then take the collinear limit $k_2=x_2P$, obtaining 
%\begin{eqnarray}
%\frac{S^{(0)}(x_2)}{x_2}\Slash S_T+S_T^\alpha S^{(1)}_{\alpha}(x_1,x_2)\Slash P
%+S_T^\alpha P^+(x_2-x_1)\left.\frac{\partial S^{(1)-}(x_1P,k_2)}
{%\partial k_2^\alpha}\right|_{k_2=x_2P}\Slash P,\label{bo2}

\subsection{$k_T$ factorization}

Next we come to the more complicated $k_T$ factorization, in which both the initial and final state partons 
can carry transverse momenta. As elaborated below, the transverse momenta $\loT$, $\ltT$ of the real gluons in 
the considered two-loop diagrams serve as these additional parton kinematic variables \cite{Nagashima:2002ia},
independent of the momentum fractions $x$ and $z$. For example, $\loT$ needs not to be equal to 
$\boldsymbol{P}_{h T}/z$ 
associated with the produced hadron as in the collinear factorization. A parton is then off-shell by $-l_T^2$ 
in the $k_T$ factorization, which is regarded as an infrared scale. That is, an infrared divergence in the 
$k_T$ factorization is represented by an infrared logarithm $\ln l_T^2$. A factorization formula is  
expressed as a convolution of a hard kernel with TMD PDFs and TMD FFs in both longitudinal 
and transverse momenta. The analysis of the phase origin is the same as in Sec.~III with the solutions of 
$l_1^+$ and $l_2^+$ being easily adapted from their collinear counterparts, given by Eqs.~(\ref{2m0}) and (\ref{2p}), 
respectively. Below we will discuss the $k_T$ factorization for the four combinations of $(l_1^+,l_2^+)$ separately.

First consider the $(++)$ case, for which the radiative gluons of the momenta $l_1$ and $l_2$ are 
both collimated to the initial proton under the hierarchy similar to Eq.~(\ref{h++}).
The two final state partons with the momenta $p_2-l_1$ and $l_1$ move in the minus and plus directions, 
respectively. The incoming quark of the momentum $p_1-l_2$ is nearly on-shell, and the associated $l_{2T}$ 
integral produces an infrared logarithm from the collinear region $l_{2T}\sim l_{1T}$ as shown in Eq.~(\ref{coll}).
Besides, the $l_2-l_1$ gluon with the invariant mass being of order $l_{1T}^2$ as $l_{2T}\sim l_{1T}$, 
is soft according to Eq.~(\ref{12}).
On the other hand, the outgoing quark of the the momentum $p_2-l_2$ moves mainly in the minus direction, namely,
in the direction of the produced hadron. Since the attaching gluon momentum $l_2-l_1$ 
is soft, the quark line with the momentum $p_2-l_2$ can be eikonalized:
\begin{eqnarray}
\frac{1}{(p_2-l_2)^2}=\frac{1}{[(p_2-l_1)+(l_1-l_2)]^2}\approx \frac{1}{2(p^-_2-l^-_1)(l_1^+-l_2^+)},
\end{eqnarray} 
if this gluon is longitudinally polarized. The resultant Wilson line contains the propagator 
$1/(l_1^+-l_2^+ +i\epsilon)$, which generates a phase as $l_1^+=l_2^+$. The collinear logarithm
together with this phase are then absorbed into the Sivers function by 
inserting the projector $(\gamma^+)(\gamma^-)$ from the Fierz identity in Eq.~(\ref{fierz0}):
the matrix $\gamma^+$ appears as the spin projector in the definition of
the Sivers function, and $\gamma^-$ is contracted to the corresponding leading-order two-parton hard kernel. 
Under this factorization, the quark carries the momentum $p_1-l_1$ before hard scattering,
implying that the Sivers function depends on the longitudinal momentum $p_1^+-l_1^+\equiv xP^+$
and the transverse momentum $l_{1T}$.
%Moreover, one can adopt a scheme, where the finite remainder after the $l_{2T}$ integration is also subtracted.   

If the $l_2-l_1$ gluon is transversely polarized, the collinear logarithm can be absorbed into the 
one-loop renormalization of the twist-three 
three-parton TMD PDF (the TMD version of the ETQS function). To achieve this factorization, we simply insert
the projector $(\gamma^+)(\gamma^-)$ from the Fierz identity: $\gamma^+$ appears as the spin 
projector in the definition of the three-parton TMD PDF, and $\gamma^-$ is contracted to the corresponding 
leading-order three-parton hard kernel. After the factorization, the quark and the gluon on the left of the 
final state cut carry the momenta $p_1-l_2$ and $l_2-l_1$ before hard scattering, respectively, and the quark 
on the right of the final state cut carries $p_1-l_1$. It indicates that the three-parton
TMD PDF depends on the longitudinal momenta $p_1^+-l_2^+\equiv x_1P^+$ and $p_1^+-l_1^+\equiv x_2P^+$ and on 
the transverse momenta $l_{1T}$ and $l_{2T}$. The phase comes from the on-shell $p_2-l_2$ propagator in the hard 
kernel, which corresponds to the SGP contribution observed in the collinear factorization as $l_{2T}=l_{1T}$, and to the 
HP contribution as $l_{2T}\not =l_{1T}$. We thus conclude that the $(++)$ component does not lead to a new 
contribution to SSA.
%, and will not be considered any longer.

Next we turn to the $(--)$ combination, for which both the radiative gluons of the momenta $l_1$ and $l_2$ 
follow the hierarchy similar to Eq.~(\ref{h--}).
Due to $p_2^-\gg p_2^+$, the two final state partons as well as the momentum $p_2-l_2$ are mainly in 
the minus direction. The incoming quark of the momentum $p_1-l_2$ is highly off-shell by ${\cal O}(Q^2)$, 
so the collinear-to-proton divergence in Eq.~(\ref{coll}) is absent.
The $l_2-l_1$ propagator develops a soft logarithm as $l_{2T}\sim l_{1T}$, the same as in the
$(++)$ combination according to Eq.~(\ref{12}). Since $l_{1,2}^+$ are soft, the two internal quark lines 
with the momenta $p_1-l_2$ and $p_1-l_1$ can be eikonalized. The resultant phase is absorbed by the 
twist-two FF, or the Collins function in the $k_T$ factorization framework.
Note that, because the eikonalized $p_1-l_{1,2}$ quark lines always remain off-shell, the Wilson lines
involved in the definition of the Collins function do not produce a phase. This result differs 
from that for the Sivers function mentioned above. See also \cite{Meissner:2008yf}.

The factorization of the infrared logarithm into the Collins FF can be done by 
inserting the Fierz identity in Eq.~(\ref{fierz0}) between the two-loop FF and the 
leading-order two-parton hard kernel (i.e., between the upper two blobs in Fig.~\ref{dia}). One picks 
up the $(\gamma_5\sigma^{i+})(\sigma_{i+}\gamma_5)$ term, in which
$\sigma_{i+}\gamma_5$ goes into the definition of the Collins  
function, and $\gamma_5\sigma^{i+}$ goes into the hard kernel. 
It implies that the same spin projector also enters the leading-order PDF of the polarized proton, 
defining the transversity distribution $h_1$. The other Dirac structures lead to either vanishing or 
subleading (twist-three TMD) contributions. The final state quark carries the momentum $p_2-l_1$, so 
the Collins function depends on the longitudinal momentum $p_2^--l_1^-\equiv zP_h^-$ and the transverse momentum 
$l_{1T}$. In conclusion, the $(--)$ contribution also reduces to the known mechanism of SSA.
%, and  we do not consider it further.

We then turn to the $(-+)$ combination. It has been pointed out that the $l_2-l_1$ propagator does not
generate an infrared logarithm in this case (see Eq.~(\ref{12})).
The quark line $p_1-l_2$ develops a collinear logarithm when the vertex $\beta$ of the $l_2$ gluon is 
transverse, as explained in the previous subsection. The $k_T$ factorization of this infrared logarithm 
is similar to the collinear factorization: it is absorbed into the three-parton TMD PDF with 
the same spin projector. Under this factorization, the quark and the gluon on the left of the 
final state cut carry the momenta $p_1-l_2$ and $l_2$ before hard scattering, respectively, and the quark 
on the right of the final state cut carries $p_1$. It indicates that the three-parton
TMD PDF depends on the longitudinal momenta $p_1^+-l_2^+\equiv x_1P^+$ and $p_1^+\equiv x_2P^+$ and on the 
transverse momentum $l_{2T}$. The phase comes from the on-shell $p_2-l_2$ propagator in the one-loop three-parton 
hard kernel, which corresponds to the HP contribution observed in the collinear factorization. There is no 
SGP contribution, because of $l_2\not= l_1$ for the $(-+)$ combination.

At last, we investigate the $(+-)$ combination, in which the phase cannot be absorbed into nonperturbative 
distribution functions. For this combination, there is no infrared singularity in the $l_2-l_1$ and 
$p_1-l_2$ propagators. The apparent singularity at $\loT=-\ltT$ from the last term of 
Eq.~(\ref{mm}) does not exist either. Hence, we arrive at a factorization formula similar to
Eq.~(\ref{our}), but with $g_T$ and $D_1$ being interpreted as the TMD PDF and the TMD FF, respectively.

%\subsection{$k_T$ factorization up to Twist 3}

Before closing this section, we briefly comment on the general structure of SSA at the two-parton 
twist-three level in the $k_T$ factorization framework. If we allow for $k_T$-dependent distributions, 
there are more contributions than the TMD versions of Eqs.~(\ref{our}) and (\ref{sup}). For example,
one can insert $(\gamma^i)(\gamma_i)$ between the upper two blobs and $(\gamma_5\gamma^+)(\gamma^-\gamma_5)$ 
between the lower two blobs. The former yields the twist-three TMD FF $D^\perp$, while the latter 
yields the twist-two TMD PDF $g_{1T}$. (All the notations for the TMD PDFs and the TMD FFs follow 
\cite{Bacchetta:2006tn}.) Exhausting all possible combinations of the spin projectors for higher-order 
hard kernels, we derive the contributions to SSA up to the two-parton twist-three
and two-loop level
\begin{eqnarray}
d\sigma&=&f^{\perp}_{1T}\otimes H_{\gamma^-,\gamma^+}^{(0)}\otimes D_1
+f^{\perp}_{1T}\otimes H_{\gamma^-,\gamma^x}^{(1)}\otimes D^{\perp}
+f^{\perp}_{1T}\otimes H_{\gamma^-,\gamma_5\gamma^x}^{(2)}\otimes G^{\perp}\nonumber\\
& &+g_{1T}\otimes H_{\gamma_5\gamma^-,\gamma^+}^{(2)}\otimes D_1
+g_{1T}\otimes H_{\gamma_5\gamma^-,\gamma_5\gamma^y}^{(1)}\otimes G^{\perp}
+g_{1T}\otimes H_{\gamma_5\gamma^-,\gamma^y}^{(2)}\otimes D^{\perp}\nonumber\\
& &+h_1\otimes H_{\gamma_5\sigma^{y-},\gamma_5\sigma^{y+}}^{(0)}\otimes H_1^{\perp}
+h_1\otimes H_{\gamma_5\sigma^{y-},\gamma_5\sigma^{yx}}^{(1)}\otimes H^*
+h_1\otimes H_{\gamma_5\sigma^{y-},I}^{(2)}\otimes E^*\nonumber\\
& &+e_T\otimes H_{\gamma_5,\gamma_5\sigma^{y+}}^{(1)}\otimes H_1^{\perp}
+e_T^\perp\otimes H_{I,\gamma_5\sigma^{y+}}^{(2)}\otimes H_1^{\perp}\nonumber\\
& &+f_{T}\otimes H_{\gamma^y,\gamma^+}^{(1)}\otimes D_1
+g_T\otimes H_{\gamma_5\gamma^y,\gamma^+}^{(2)}\otimes D_1\nonumber\\
& &+h_T^\perp\otimes H_{\gamma_5\sigma^{yx},\gamma_5\sigma^{y+}}^{(1)}\otimes H_1^{\perp}
+h_T\otimes H_{\gamma_5\sigma^{-+},\gamma_5\sigma^{y+}}^{(1)}\otimes H_1^{\perp}, \label{many}
\end{eqnarray}
where the functions labelled by $*$ diminish for a massless produced hadron.
The FF $G^\perp$ comes from the projector $\gamma_i\gamma_5$, and $H_1^\perp$ from $\sigma_{i+}\gamma_5$.
The TMD transversity function $h_1$ denotes $h_1-(k_x^2-k_y^2)h_{1T}^\perp/(2M^2)$ actually. For the 
$h_1$ piece, the hard kernel $H^{(1)}_{\gamma_5\sigma^{y-},\gamma_5\sigma^{x+}}$ may appear
at one loop. It has been omitted in Eq.~(\ref{many}), because it is subleading compared to the term 
$H^{(0)}_{\gamma_5\sigma^{y-},\gamma_5\sigma^{y+}}$. The nonperturbative spin-momentum correlation in 
the Sivers function and the Collins function are basically determined by fits to
data. Including the numerous terms in Eq.~(\ref{many}), it is expected
to make an impact on the determination of the Sivers function and the Collins function.

When we work in the collinear factorization, all the above terms  vanish except for the ones which reduce 
to Eqs.~(\ref{our}) and (\ref{sup}). This emphasizes the importance of the parton transverse momentum 
for the existence of SSA. Among the many terms in Eq.~(\ref{many}), the one proportional to the 
distribution $f_T$ is particularly interesting. Since $f_T$ is T-odd, the corresponding contribution flips 
signs between SIDIS and Drell-Yan. Its definition involves the proton spin 
$\langle \bar{\psi}\gamma^\alpha\psi\rangle \sim \epsilon_T^{\alpha\beta}S_{T\beta} f_T(x,k_T^2)$, that 
combines with a factor of $k^x$ from the one-loop hard kernel $H^{(1)}$ to generate a SSA
proportional to $P_h^xS_T^y$. If we stick to the leading order hard kernel, the $k^x$ dependence
will disappear, and $f_T$ will contribute only to the SIDIS structure function associated with 
$\sin \phi_S$ (denoted by $F_{UT}^{\sin \phi_S}$ in \cite{Bacchetta:2006tn}), where
$\phi_S$ is the azimuthal angle of the proton spin relative to the lepton plane.
Because the first moment vanishes $\int d^2k_T f_T(x,k_T^2)=0$,
its $k_T$ dependence exhibits some nodes in $k_T$.
This may result in a node in SSA as a function of $P_{hT}$, similarly to what was observed in 
\cite{Zhou:2013gsa,Yao:2018vcg}.

\section{Conclusion}

In this paper, we have presented a detailed study of the two-loop diagrams that produce an imaginary phase in SIDIS and discussed their gauge invariance and collinear factorization properties. In addition to the known mechanisms for SSA, we have also identified an entirely new contribution proportional to the $g_T$ distribution function. While it is parametrically suppressed by a factor $\alpha_s$, $g_T$ has the Wandzura-Wilczek part related to the polarized quark distribution functions. Since this part is usually considered to be larger than the genuine twist-three one, our new contribution could be comparable in magnitude to those from the ETQS function. In a future publication \cite{prep}, we plan to give a numerical estimate of the obtained results in this paper, and  make comparisons with the existing data as well as predictions for the Electron-Ion Collider.

We note that there have been a lot of discussions on potentially dominant sources of SSA recently. There is an indication that the Sivers or ETQS contribution may be numerically small \cite{Kang:2011hk}. Instead, a successful fit of the RHIC data \cite{Kanazawa:2014dca,Gamberg:2017gle} suggests that the twist-three FFs may be the dominant source of SSA. In order to confirm this, the same FFs should be able to fit other observables \cite{Hatta:2016khv,Zhou:2017sdx,Benic:2018moa,Benic:2018amn}. In other words, a global analysis of many different data is necessary for understanding the above observations. The subleading contributions derived in the $k_T$ factorization with a more complete set of origins for SSA may provide such a theoretical framework.
Because the momentum transferred involved in the relevant processes are not large enough, higher-order
hard cross sections may give sizable corrections. Therefore,
the rich subleading structures proposed in this work are phenomenologically important.

\begin{acknowledgments}

%\vskip 0.2cm
S.~B. and H.~L. thank the nuclear theory group of Brookhaven National Laboratory for support and hospitality.  
H.~L. and D.~Y. thank Yukawa Institute for Theoretical Physics, Kyoto University for hospitality. We thank 
Zhong-Bo Kang, Yuji Koike, Marc Schlegel, Werner Vogelsang,  Shinsuke Yoshida and Jian Zhou for useful discussions. 
This material is based upon work supported by  the U.S. Department of Energy, 
Office of Science, Office of Nuclear Physics, under contract number  DE-SC0012704  and the LDRD program of Brookhaven 
National Laboratory. It is also in part supported by the Ministry of Science and Technology of R.O.C. under
Grant No. MOST-107-2119-M-001-035-MY3. S. B. is supported by a JSPS
postdoctoral fellowship for foreign researchers under Grant No. 17F17323.

\end{acknowledgments}


\begin{thebibliography}{99}



\bibitem{B76} G. Bunce et al., Phys. Rev. Lett. {\bf 36}, 1113 (1976);
K. Heller et al., Phys. Lett. B {\bf 68}, 480 (1977); S.A. Gourlay et al.,
Phys. Rev. Lett. {\bf 56}, 2244 (1986).

\bibitem{A91} D.L. Adams et al., Phys. Lett. B {\bf 261}, 201 (1991);
B {\bf 264}, 462 (1991); A. Bravar et al., Phys. Rev. Lett. {\bf 77}, 2626 (1996).

%\cite{Adare:2013ekj}
\bibitem{Adare:2013ekj}
  A.~Adare {\it et al.} [PHENIX Collaboration],
  %``Measurement of transverse-single-spin asymmetries for midrapidity and forward-rapidity production of hadrons in polarized p+p collisions at $\sqrt{s}=$200 and 62.4 GeV,''
  Phys.\ Rev.\ D {\bf 90}, 012006 (2014).
  %doi:10.1103/PhysRevD.90.012006
  %[arXiv:1312.1995 [hep-ex]].
  %%CITATION = doi:10.1103/PhysRevD.90.012006;%%
  %73 citations counted in INSPIRE as of 27 Jul 2

%\cite{Adamczyk:2012xd}
\bibitem{Adamczyk:2012xd}
  L.~Adamczyk {\it et al.} [STAR Collaboration],
  %``Transverse Single-Spin Asymmetry and Cross-Section for $\pi^0$ and $\eta$ Mesons at Large Feynman-$x$ in Polarized $p+p$ Collisions at $\sqrt{s}=200$ GeV,''
  Phys.\ Rev.\ D {\bf 86}, 051101 (2012)
  %doi:10.1103/PhysRevD.86.051101
  %[arXiv:1205.6826 [nucl-ex]].
  %%CITATION = doi:10.1103/PhysRevD.86.051101;%%
  %75 citations counted in INSPIRE as of 27 Jul 2019


\bibitem{KPR78} G.L. Kane, J. Pumplin and W. Repko, Phys. Rev. Lett.
{\bf 41}, 1689 (1978).


%\cite{Efremov:1981sh}
\bibitem{Efremov:1981sh}
  A.~V.~Efremov and O.~V.~Teryaev,
  %``On Spin Effects in Quantum Chromodynamics,''
  Sov.\ J.\ Nucl.\ Phys.\  {\bf 36}, 140 (1982)
  [Yad.\ Fiz.\  {\bf 36}, 242 (1982)].
  %%CITATION = SJNCA,36,140;%%
  %306 citations counted in INSPIRE as of 27 Jul 2019


%\cite{Efremov:1983eb}
\bibitem{Efremov:1983eb}
  A.~V.~Efremov and O.~V.~Teryaev,
  %``The Transversal Polarization In Quantum Chromodynamics,''
  Sov.\ J.\ Nucl.\ Phys.\  {\bf 39}, 962 (1984)
  [Yad.\ Fiz.\  {\bf 39}, 1517 (1984)].
  %%CITATION = SJNCA,39,962;%%
  %114 citations counted in INSPIRE as of 27

%\cite{Efremov:1984ip}
\bibitem{Efremov:1984ip}
  A.~V.~Efremov and O.~V.~Teryaev,
  %``QCD Asymmetry and Polarized Hadron Structure Functions,''
  Phys.\ Lett.\  {\bf 150B}, 383 (1985).
  %doi:10.1016/0370-2693(85)90999-2
  %%CITATION = doi:10.1016/0370-2693(85)90999-2;%%
  %337 citations counted in INSPIRE as of 27 Jul 2019

%\cite{Ratcliffe:1985mp}
\bibitem{Ratcliffe:1985mp}
  P.~G.~Ratcliffe,
  %``Transverse Spin and Higher Twist in {QCD},''
  Nucl.\ Phys.\ B {\bf 264}, 493 (1986).
  %doi:10.1016/0550-3213(86)90495-5
  %%CITATION = doi:10.1016/0550-3213(86)90495-5;%%
  %117 citations counted in INSPIRE as of 27 Jul 2019



%\cite{Qiu:1991pp}
\bibitem{Qiu:1991pp}
  J.~w.~Qiu and G.~F.~Sterman,
  %``Single transverse spin asymmetries,''
  Phys.\ Rev.\ Lett.\  {\bf 67}, 2264 (1991).
  %doi:10.1103/PhysRevLett.67.2264
  %%CITATION = doi:10.1103/PhysRevLett.67.2264;%%
  %470 citations counted in INSPIRE as of 27 Jul 2019

%\cite{Qiu:1998ia}
\bibitem{Qiu:1998ia}
  J.~w.~Qiu and G.~F.~Sterman,
  %``Single transverse spin asymmetries in hadronic pion production,''
  Phys.\ Rev.\ D {\bf 59}, 014004 (1999).
  %doi:10.1103/PhysRevD.59.014004
  %[hep-ph/9806356].
  %%CITATION = doi:10.1103/PhysRevD.59.014004;%%
  %416 citations counted in INSPIRE as of 27 J


%\cite{Kang:2010zzb}
\bibitem{Kang:2010zzb}
  Z.~B.~Kang, F.~Yuan and J.~Zhou,
  %``Twist-three fragmentation function contribution to the single spin asymmetry in p p collisions,''
  Phys.\ Lett.\ B {\bf 691}, 243 (2010).
  %doi:10.1016/j.physletb.2010.07.003
  %[arXiv:1002.0399 [hep-ph]].
  %%CITATION = doi:10.1016/j.physletb.2010.07.003;%%
  %82 citations counted in INSPIRE as of



%\cite{Metz:2012ct}
\bibitem{Metz:2012ct}
  A.~Metz and D.~Pitonyak,
  %``Fragmentation contribution to the transverse single-spin asymmetry in proton-proton collisions,''
  Phys.\ Lett.\ B {\bf 723}, 365 (2013);
  Erratum: [Phys.\ Lett.\ B {\bf 762}, 549 (2016)].
  %doi:10.1016/j.physletb.2013.05.043, 10.1016/j.physletb.2016.10.011
  %[arXiv:1212.5037 [hep-ph]].
  %%CITATION = doi:10.1016/j.physletb.2013.05.043, 10.1016/j.physletb.2016.10.011;%%
  %69 citations counted in INSPIRE as


%\cite{Kanazawa:2013uia}
\bibitem{Kanazawa:2013uia}
  K.~Kanazawa and Y.~Koike,
  %``Contribution of twist-3 fragmentation function to single transverse-spin asymmetry in semi-inclusive deep inelastic scattering,''
  Phys.\ Rev.\ D {\bf 88}, 074022 (2013).
  %doi:10.1103/PhysRevD.88.074022
  %[arXiv:1309.1215 [hep-ph]].
  %%CITATION = doi:10.1103/PhysRevD.88.074022;%%
  %37 citations counted in INSPIRE



\bibitem{S90} D. Sivers, Phys. Rev. D {\bf 41}, 83 (1990); Phys. Rev.,
D {\bf 43}, 261 (1991).

\bibitem{ABM95} M. Anselmino, M. Boglione and F. Murgia, Phys. Lett.
B {\bf 362}, 164 (1995).

\bibitem{C93} J. Collins, Nucl. Phys. B {\bf 396}, 161 (1993).

%\cite{Collins:1993kq}
\bibitem{Collins:1993kq}
  J.~C.~Collins, S.~F.~Heppelmann and G.~A.~Ladinsky,
  %``Measuring transversity densities in singly polarized hadron hadron and lepton - hadron collisions,''
  Nucl.\ Phys.\ B {\bf 420}, 565 (1994).
  %doi:10.1016/0550-3213(94)90078-7
  %[hep-ph/9305309].
  %%CITATION = doi:10.1016/0550-3213(94)90078-7;%%
  %259 citations counted in INSPIRE as

\bibitem{ACY97} X. Artru , J. Czyzewski and H. Yabuki, Z. Phys. C {\bf 73},
527 (1997).

%\cite{Ma:2008gm}
\bibitem{Ma:2008gm} 
  J.~P.~Ma and H.~Z.~Sang,
  %``Partonic State and Single Transverse Spin Asymmetry in Drell-Yan Process,''
  JHEP {\bf 0811}, 090 (2008).
  %doi:10.1088/1126-6708/2008/11/090
  %[arXiv:0809.4811 [hep-ph]].
  %%CITATION = doi:10.1088/1126-6708/2008/11/090;%%
  %15 citations counted in INSPIRE as of 02 Sep 2019


%\cite{Ma:2008cj}
\bibitem{Ma:2008cj} 
  J.~P.~Ma and H.~Z.~Sang,
  %``Partonic State and Single Transverse Spin Asymmetries in Semi Inclusive DIS,''
  Phys.\ Lett.\ B {\bf 676}, 74 (2009).
  %doi:10.1016/j.physletb.2009.04.071
  %[arXiv:0811.0224 [hep-ph]].
  %%CITATION = doi:10.1016/j.physletb.2009.04.071;%%
  %11 citations counted in INSPIRE as of 02 Sep 2019

%\cite{Metz:2006pe}
\bibitem{Metz:2006pe}
  A.~Metz, M.~Schlegel and K.~Goeke,
  %``Transverse single spin asymmetries in inclusive deep-inelastic scattering,''
  Phys.\ Lett.\ B {\bf 643}, 319 (2006).
  %doi:10.1016/j.physletb.2006.11.009
  %[hep-ph/0610112].
  %%CITATION = doi:10.1016/j.physletb.2006.11.009;%%
  %22 citations counted in INSPIRE as of 27 Jul 2019


  %\cite{Afanasev:2007ii}
\bibitem{Afanasev:2007ii}
  A.~Afanasev, M.~Strikman and C.~Weiss,
  %``Transverse target spin asymmetry in inclusive DIS with two-photon exchange,''
  Phys.\ Rev.\ D {\bf 77}, 014028 (2008).
  %doi:10.1103/PhysRevD.77.014028
  %[arXiv:0709.0901 [hep-ph]].
  %%CITATION = doi:10.1103/PhysRevD.77.014028;%%
  %32 citations counted in INSPIRE as of 08 Aug 2019



\bibitem{Schlegel:2012ve} 
  M.~Schlegel,
  %``Partonic description of the transverse target single-spin asymmetry in inclusive deep-inelastic scattering,''
  Phys.\ Rev.\ D {\bf 87}, 034006 (2013).
  %doi:10.1103/PhysRevD.87.034006
  %[arXiv:1211.3579 [hep-ph]].

\bibitem{Bacchetta:2008xw}
  A.~Bacchetta, D.~Boer, M.~Diehl and P.~J.~Mulders,
  %``Matches and mismatches in the descriptions of semi-inclusive processes at low and high transverse momentum,''
  JHEP {\bf 0808}, 023 (2008).
%  doi:10.1088/1126-6708/2008/08/023
%  [arXiv:0803.0227 [hep-ph]].
  %%CITATION = doi:10.1088/1126-6708/2008/08/023;%%
  %171 citations counted in INSPIRE as of 10 Aug 2017




 %\cite{Vogelsang:2009pj}
\bibitem{Vogelsang:2009pj} 
  W.~Vogelsang and F.~Yuan,
  %``Next-to-leading Order Calculation of the Single Transverse Spin Asymmetry in the Drell-Yan Process,''
  Phys.\ Rev.\ D {\bf 79}, 094010 (2009).
  %doi:10.1103/PhysRevD.79.094010
  %[arXiv:0904.0410 [hep-ph]].
  %%CITATION = doi:10.1103/PhysRevD.79.094010;%%
  %85 citations counted in INSPIRE as of 02 Sep 2019


\bibitem{Song:2010pf}
  Y.~k.~Song, J.~h.~Gao, Z.~t.~Liang and X.~N.~Wang,
  %``Twist-4 contributions to the azimuthal asymmetry in SIDIS,''
  Phys.\ Rev.\ D {\bf 83}, 054010 (2011).
%  doi:10.1103/PhysRevD.83.054010
%  [arXiv:1012.4179 [hep-ph]].
  %%CITATION = doi:10.1103/PhysRevD.83.054010;%%
  %22 citations counted in INSPIRE as of 10 Aug 2017

%\cite{Kang:2012ns}
\bibitem{Kang:2012ns} 
  Z.~B.~Kang, I.~Vitev and H.~Xing,
  %``Transverse momentum-weighted Sivers asymmetry in semi-inclusive deep inelastic scattering at next-to-leading order,''
  Phys.\ Rev.\ D {\bf 87}, 034024 (2013).
  %doi:10.1103/PhysRevD.87.034024
  %[arXiv:1212.1221 [hep-ph]].
  %%CITATION = doi:10.1103/PhysRevD.87.034024;%%
  %35 citations counted in 






%\cite{Yoshida:2016tfh}
\bibitem{Yoshida:2016tfh} 
  S.~Yoshida,
  %``New pole contribution to $P_{h\perp}$-weighted single-transverse spin asymmetry in semi-inclusive deep inelastic scattering,''
  Phys.\ Rev.\ D {\bf 93}, 054048 (2016).
  %doi:10.1103/PhysRevD.93.054048
  %[arXiv:1601.07737 [hep-ph]].
  %%CITATION = doi:10.1103/PhysRevD.93.054048;%%
  %9 citations counted in INSPIRE as of 02 Se

%\cite{Chen:2017lvx}
\bibitem{Chen:2017lvx} 
  A.~P.~Chen, J.~P.~Ma and G.~P.~Zhang,
  %``One-Loop Corrections of Single Spin Asymmetries in Semi-Inclusive DIS,''
  Phys.\ Rev.\ D {\bf 97}, 054003 (2018).
  %doi:10.1103/PhysRevD.97.054003
  %[arXiv:1708.09091 [hep-ph]].
  %%CITATION = doi:10.1103/PhysRevD.97.054003;%%
  %3 citations counted in INSPIRE as of 02 Sep 2019


%\cite{Eguchi:2006mc}
\bibitem{Eguchi:2006mc}
  H.~Eguchi, Y.~Koike and K.~Tanaka,
  %``Twist-3 Formalism for Single Transverse Spin Asymmetry Reexamined: Semi-Inclusive Deep Inelastic Scattering,''
  Nucl.\ Phys.\ B {\bf 763}, 198 (2007).
  %doi:10.1016/j.nuclphysb.2006.11.016
  %[hep-ph/0610314].
  %%CITATION = doi:10.1016/j.nuclphysb.2006.11.016;%%
  %126 citations counted in INSPIRE as





%\cite{Hatta:2013wsa}
\bibitem{Hatta:2013wsa}
  Y.~Hatta, K.~Kanazawa and S.~Yoshida,
  %``Double-spin asymmetry $A_{LT}$ in open charm production,''
  Phys.\ Rev.\ D {\bf 88}, 014037 (2013).
  %doi:10.1103/PhysRevD.88.014037
  %[arXiv:1305.7001 [hep-ph]].
  %%CITATION = doi:10.1103/PhysRevD.88.014037;%%
  %12 citations counted in INSPIRE as


%\cite{Xing:2019ovj}
\bibitem{Xing:2019ovj}
  H.~Xing and S.~Yoshida,
  %``New approach to the Sivers effect in the collinear twist-3 formalism,''
  arXiv:1904.02287 [hep-ph].
  %%CITATION = ARXIV:1904.02287;%%
  
%\cite{Kanazawa:2015ajw}
\bibitem{Kanazawa:2015ajw} 
  K.~Kanazawa, Y.~Koike, A.~Metz, D.~Pitonyak and M.~Schlegel,
  %``Operator Constraints for Twist-3 Functions and Lorentz Invariance Properties of Twist-3 Observables,''
  Phys.\ Rev.\ D {\bf 93}, 054024 (2016).
  %doi:10.1103/PhysRevD.93.054024
  %[arXiv:1512.07233 [hep-ph]].
  %%CITATION = doi:10.1103/PhysRevD.93.054024;%%
  %27 citations counted in INSPIRE as of 03 Sep 2019


%\cite{Eguchi:2006qz}
\bibitem{Eguchi:2006qz}
  H.~Eguchi, Y.~Koike and K.~Tanaka,
  %``Single Transverse Spin Asymmetry for Large-p(T) Pion Production in Semi-Inclusive Deep Inelastic Scattering,''
  Nucl.\ Phys.\ B {\bf 752}, 1 (2006).
  %doi:10.1016/j.nuclphysb.2006.05.036
  %[hep-ph/0604003].
  %%CITATION = doi:10.1016/j.nuclphysb.2006.05.036;%%
  %108 citations counted in INSPIRE as o


%\cite{Jaffe:1991ra}
\bibitem{Jaffe:1991ra}
  R.~L.~Jaffe and X.~D.~Ji,
  %``Chiral odd parton distributions and Drell-Yan processes,''
  Nucl.\ Phys.\ B {\bf 375}, 527 (1992).
  %doi:10.1016/0550-3213(92)90110-W
  %%CITATION = doi:10.1016/0550-3213(92)90110-W;%%
  %611 citations counted in INSPIRE a

\bibitem{BS89} J.C. Collin and D.E. Soper, Nucl. Phys. B {\bf 193}, 381 (1981);
J. Botts and G. Sterman, Nucl. Phys. B {\bf 325}, 62 (1989);
H. n. Li and H.L. Yu, Phys. Rev. Lett. {\bf 74}, 4388 (1995).

\bibitem{Brodsky:2002cx} 
  S.~J.~Brodsky, D.~S.~Hwang and I.~Schmidt,
  %``Final state interactions and single spin asymmetries in semiinclusive deep inelastic scattering,''
  Phys.\ Lett.\ B {\bf 530}, 99 (2002).
  %doi:10.1016/S0370-2693(02)01320-5
  %[hep-ph/0201296].




\bibitem{Bacchetta:2006tn}
  A.~Bacchetta, M.~Diehl, K.~Goeke, A.~Metz, P.~J.~Mulders and M.~Schlegel,
  %``Semi-inclusive deep inelastic scattering at small transverse momentum,''
  JHEP {\bf 0702}, 093 (2007).
%  doi:10.1088/1126-6708/2007/02/093
%  [hep-ph/0611265].
  %%CITATION = doi:10.1088/1126-6708/2007/02/093;%%
  %455 citations counted in INSPIRE as of 10 Aug 2017






%\cite{Wandzura:1977qf}
\bibitem{Wandzura:1977qf}
  S.~Wandzura and F.~Wilczek,
  %``Sum Rules for Spin Dependent Electroproduction: Test of Relativistic Constituent Quarks,''
  Phys.\ Lett.\  {\bf 72B}, 195 (1977).
  %doi:10.1016/0370-2693(77)90700-6
  %%CITATION = doi:10.1016/0370-2693(77)90700-6;%%
  %508 citations counted in INSPIRE as of 28



\bibitem{Kanazawa:2014dca}
  K.~Kanazawa, Y.~Koike, A.~Metz and D.~Pitonyak,
  %``Towards an explanation of transverse single-spin asymmetries in proton-proton collisions: the role of fragmentation in collinear factorization,''
  Phys.\ Rev.\ D {\bf 89}, 111501 (2014).
%  doi:10.1103/PhysRevD.89.111501
%  [arXiv:1404.1033 [hep-ph]].

\bibitem{prep}
S.~Beni\' c, Y.~Hatta, H.~n.~Li and D.~J.~ Yang, in progress.



%\cite{Kang:2008ey}
\bibitem{Kang:2008ey} 
  Z.~B.~Kang and J.~W.~Qiu,
  %``Evolution of twist-3 multi-parton correlation functions relevant to single transverse-spin asymmetry,''
  Phys.\ Rev.\ D {\bf 79}, 016003 (2009).
  %doi:10.1103/PhysRevD.79.016003
  %[arXiv:0811.3101 [hep-ph]].
  %%CITATION = doi:10.1103/PhysRevD.79.016003;%%
  %97 citations counted in INSPIRE as of 19 Sep 2019


\bibitem{Nagashima:2002ia} 
  M.~Nagashima and H.~n.~Li,
  %``k(T) factorization of exclusive processes,''
  Phys.\ Rev.\ D {\bf 67}, 034001 (2003).



%\cite{Meissner:2008yf}
\bibitem{Meissner:2008yf}
  S.~Meissner and A.~Metz,
  %``Partonic pole matrix elements for fragmentation,''
  Phys.\ Rev.\ Lett.\  {\bf 102}, 172003 (2009).
  %doi:10.1103/PhysRevLett.102.172003
  %[arXiv:0812.3783 [hep-ph]].
  %%CITATION = doi:10.1103/PhysRevLett.102.172003;%%
  %63 citations counted in INSPIRE as of 05 Jul 2019

%\cite{Zhou:2013gsa}
\bibitem{Zhou:2013gsa} 
  J.~Zhou,
  %``Transverse single spin asymmetries at small x and the anomalous magnetic moment,''
  Phys.\ Rev.\ D {\bf 89}, 074050 (2014).
  %doi:10.1103/PhysRevD.89.074050
  %[arXiv:1308.5912 [hep-ph]].

%\cite{Yao:2018vcg}
\bibitem{Yao:2018vcg}
  X.~Yao, Y.~Hagiwara and Y.~Hatta,
  %``Computing the gluon Sivers function at small-$x$,''
  Phys.\ Lett.\ B {\bf 790}, 361 (2019).
  %doi:10.1016/j.physletb.2019.01.029
  %[arXiv:1812.03959 [hep-ph]].
  %%CITATION = doi:10.1016/j.physletb.2019.01.029;%%
  %1 citations counted in INSPIRE as of 06 Jul 2019




\bibitem{Kang:2011hk}
  Z.~B.~Kang, J.~W.~Qiu, W.~Vogelsang and F.~Yuan,
  %``An Observation Concerning the Process Dependence of the Sivers Functions,''
  Phys.\ Rev.\ D {\bf 83}, 094001 (2011).
%  doi:10.1103/PhysRevD.83.094001
%  [arXiv:1103.1591 [hep-ph]].






\bibitem{Gamberg:2017gle}
  L.~Gamberg, Z.~B.~Kang, D.~Pitonyak and A.~Prokudin,
  %``Phenomenological constraints on $A_N$ in $p^\uparrow p\to \pi\, X$ from Lorentz invariance relations,''
  Phys.\ Lett.\ B {\bf 770}, 242 (2017).
%  doi:10.1016/j.physletb.2017.04.061
%  [arXiv:1701.09170 [hep-ph]].






\bibitem{Hatta:2016khv}
  Y.~Hatta, B.~W.~Xiao, S.~Yoshida and F.~Yuan,
  %``Single spin asymmetry in forward $pA$ collisions II: Fragmentation contribution,''
  Phys.\ Rev.\ D {\bf 95}, 014008 (2017).
%  doi:10.1103/PhysRevD.95.014008
%  [arXiv:1611.04746 [hep-ph]].

\bibitem{Zhou:2017sdx}
  J.~Zhou,
  %``Single spin asymmetries in forward p-p/A collisions revisited: the role of color entanglement,''
  Phys.\ Rev.\ D {\bf 96}, 034027 (2017).
%  doi:10.1103/PhysRevD.96.034027
%  [arXiv:1704.04901 [hep-ph]].

%\cite{Benic:2018moa}
\bibitem{Benic:2018moa}
  S.~Benić and Y.~Hatta,
  %``Single spin asymmetries in ultra-peripheral $p^\uparrow A$ collisions,''
  Phys.\ Rev.\ D {\bf 98}, 094025 (2018).
  %doi:10.1103/PhysRevD.98.094025
  %[arXiv:1806.10901 [hep-ph]].
  %%CITATION = doi:10.1103/

%\cite{Benic:2018amn}
\bibitem{Benic:2018amn}
  S.~Benić and Y.~Hatta,
  %``Single spin asymmetry in forward $pA$ collisions: Phenomenology at RHIC,''
  Phys.\ Rev.\ D {\bf 99}, 094012 (2019).
  %doi:10.1103/PhysRevD.99.094012
  %[arXiv:1811.10589 [hep-ph]].
  %%CITATION = doi:10.1103/PhysRevD.99.09




\end{thebibliography}
\end{document}